\documentclass[preprint,
superscriptaddress,
 amsmath,amssymb,
 aps,
]{revtex4-2}

\usepackage{amsmath}
\usepackage{amssymb}
\usepackage{epsfig}
\usepackage{esint}
\usepackage{xcolor}
\usepackage{tikz}
\usepackage{appendix}
\usepackage{graphicx}
\usepackage{subcaption}
\usepackage{comment}
\usepackage{ulem}
\usepackage{mathtools}

\usepackage[greek,english]{babel}
\usepackage{teubner}

\newcommand{\br}{{\bf r}}

\newcommand{\rhat}{{\bf \hat r}}

\newcommand{\by}{{\bf y}}

\newcommand{\ehat}{{\bf{\hat e}}}
\newcommand{\that}{{\bm{\hat \theta}}}
\newcommand{\phihat}{{\bm{\hat \varphi}}}

\newcommand{\bv}{{\bf v}}

\newcommand{\im}{{\mathrm i}}

\newcommand{\half} {{\frac{1}{2}}}

\newcommand{\bS}{{\bf{y}}}

\definecolor{mathematica_blue}{RGB}{175,192,218}
\definecolor{mathematica_yellow}{RGB}{253,214,148}
\definecolor{mathematica_green}{RGB}{159,184,133}

\begin{document}

\normalem
\pagenumbering{gobble}

\title{A robust method for quantification of \\ surface elasticity in soft solids}

\author{Stefanie Heyden}
 \email{stefanie.heyden@mat.ethz.ch}
 \affiliation{Department of Materials, ETH Z\"{u}rich, 8093 Z\"{u}rich, Switzerland.}
 
\author{Petia M. Vlahovska}
 \email{petia.vlahovska@northwestern.edu}
 \affiliation{Engineering Sciences and Applied Mathematics, Northwestern University, Evanston, Illinois 60208, USA.}

\author{Eric R. Dufresne}
 \email{eric.dufresne@mat.ethz.ch}
 \affiliation{Department of Materials, ETH Z\"{u}rich, 8093 Z\"{u}rich, Switzerland.}

\begin{abstract}
We propose an approach to measure  surface elastic constants of soft solids. Generally, this requires one to probe interfacial mechanics at around the elastocapillary length scale, which is typically microscopic.
Deformations of  microscopic droplets embedded in soft solids are particularly attractive, because they avoid intrinsic nonlinearities associated with previous experiments such as the equilibrium of contact lines and the relaxation of patterned surfaces.
We derive analytical solutions for the shape of droplets under uniaxial deformation and for the radius of droplets upon hydrostatic inflation.
We couple mechanical deformations to the dissolution of droplets to assess experimental sensitivities.
Combined with experimental data from both modes of deformation, one should be able to reliably extract the complete set of isotropic surface material parameters following a specific minimization procedure.
\end{abstract}

\maketitle

\section{Introduction}

\textLipsias{>epif'an-eia}, [\textLipsias{<h}]: \textit{visible surface} of a body.
In ancient greek, the linguistic distinction between an object's boundary and interior dates back to the centuries \emph{a.Chr.n.} \cite{Liddell:1843}.
It took another two millenia for the corresponding concepts to be laid out in fluid- and solid mechanics \cite{Pockels:1891,Pockels:1898,Pockels:1899,Rayleigh:1916,Shuttleworth:1950,Gurtin:1975,Gurtin:1998}. At the tip of this timeline, the debate on surface mechanics of soft solids is ongoing \cite{Dervaux:2015,Jensen:2017,Xu:2017,Style:2017,Xu:2018,Style:2018,Masurel:2019,Heyden:2021}. 
The controversy around surface elastic effects in soft solids is rooted in their hybrid nature, incorporating features of both fluids and solids. 
While model systems such as polymer networks exhibit fluid-like behavior at scales below the polymer mesh size, the presence of crosslinks gives rise to a finite shear modulus at the macroscale as in elastic solids \cite{Pandey:2020}. 

Experimental evidence for surface elasticity of gels has emerged in a variety of settings. 
The coupling of elasticity to capillarity was first examined in \cite{Mora:2010}.
Further studies include static wetting on strained substrates \cite{Xu:2017,Xu:2018}, soft solid contacts \cite{Jensen:2017}, and the strain-dependent topography of soft solids \cite{Bain:2021}. 
In all settings, deformations are induced at the scale of the elastocapillary length, which is on the order of micrometers in gels (E $\sim\mathcal{O}$(kPa)). 
At this scale, surface effects start to dominate over the bulk behavior.

Previous theoretical studies on surface mechanical properties in solids are based on the pioneering work of Gurtin and Murdoch \cite{Gurtin:1975,Gurtin:1998}. 
Investigations have focused on point- or line-forces acting upon an elastic substrate as induced by static wetting (see, \emph{e.g.}, \cite{Jerison:2011,Style:2012,Lubarda:2013,Bostwick:2014,Sauer:2014,Dervaux:2015,Pandey:2020,Heyden:2021}). 
Additional studies have examined surface-tension-induced flattening of nearly plane elastic solids \cite{Jagota:2012}, the strain-dependent topography of soft solids \cite{Bain:2021}, as well as the effects of capillarity in inhomogeneous soft materials on Euler buckling \cite{Wettlaufer:2021}. 
Extensions to the canonical problem of an inclusion embedded in an elastic matrix (known as \emph{Eshelby's inclusion problem}) have furthermore accounted for surface elastic effects at the interface (cf., \emph{e.g.}, \cite{Duan:2005,Style:20152,Shiavone:2018,Sharma:2020}).
A recent work \cite{Basu:2021} extends theoretical investigations to the calibration of surface hyperelastic models in soft solids through a combination of tensile and torsion tests on slender objects.

However, several drawbacks can impede a clean interpretation of experimental data: Attendant large deformations (\emph{i.e.} via applied pre-stretches \cite{Xu:2017,Xu:2018} or initial relaxations of patterned surfaces \cite{Bain:2021}) can induce material- and geometric nonlinearities. 
In addition, geometric singularities such as the tip of a wetting ridge cause stress divergences within the bulk material \cite{Jerison:2011,Masurel:2019,Pandey:2020}.

Here, we propose a series of mechanical tests to reliably quantify surface elasticity in soft solids. 
For isotropic, linear-elastic surfaces, three surface material parameters need to be determined: the residual surface stress $\Upsilon^0$, as well as surface shear- and bulk moduli $\mu^s$ and $\kappa^s$, respectively.
We consider two distinct deformation modes: 1) The shape of droplets embedded in an elastic matrix under uniaxial strain, and 2) The dilation of condensing droplets under hydrostatic pressure within an elastic matrix. 
Both modes of deformation presented here feature geometrically smooth surfaces, hence circumventing geometric singularities. 
In addition, the proposed deformation modes are accessible in the small deformation regime.
The reference configuration is that of a smooth undisturbed surface, hence possible nonlinearities are eliminated. 
Inspired by the dual nature of soft solids, we utilize methods from both fluid- and solid mechanics to solve for interface deformation upon loading.
The proposed data analysis allows for a reliable recovery of the complete set of surface parameters $(\Upsilon^0,\mu^s,\kappa^s)$. 
In addition, fitting sensitivities are estimated for all physically admissible classes of parameter ranges. 
We find that a robust extraction of relevant quantities is achieved for good experimental signal to noise ratios.

\section{Gedankenexperimente}
\label{sec:Experiments}

Consider a dispersion of microscopic liquid droplets in a soft solid (see Figure 1).  
This can be fabricated in two  steps \cite{Style:20152}. First, the liquid is  dispersed  in a  polymeric precursor at a temperature, $T^{mix}$.
The  molecules making up the droplet phase partially dissolve to reach an equilibrium concentration in the solid phase, $n_{sat}(T^{mix})$.
Then, the sample is crosslinked,  and  the droplets are surrounded by a stress-free solid network.
At sufficient dilution, each droplet will respond independently to imposed stresses.

We propose two complementary mechanical probes of these samples: macroscopic uniaxial deformation and microscopic inflation.
As we will see, these two tests are sensitive to different surface mechanical parameters.
The smooth surfaces of droplets avoids nonlinear stress concentrations, and the tests can be conducted in the small deformation limit.

\begin{figure}[ht]
\begin{tikzpicture}
    \node (exp_sym) at (0,0) {\includegraphics[width=0.85\textwidth]{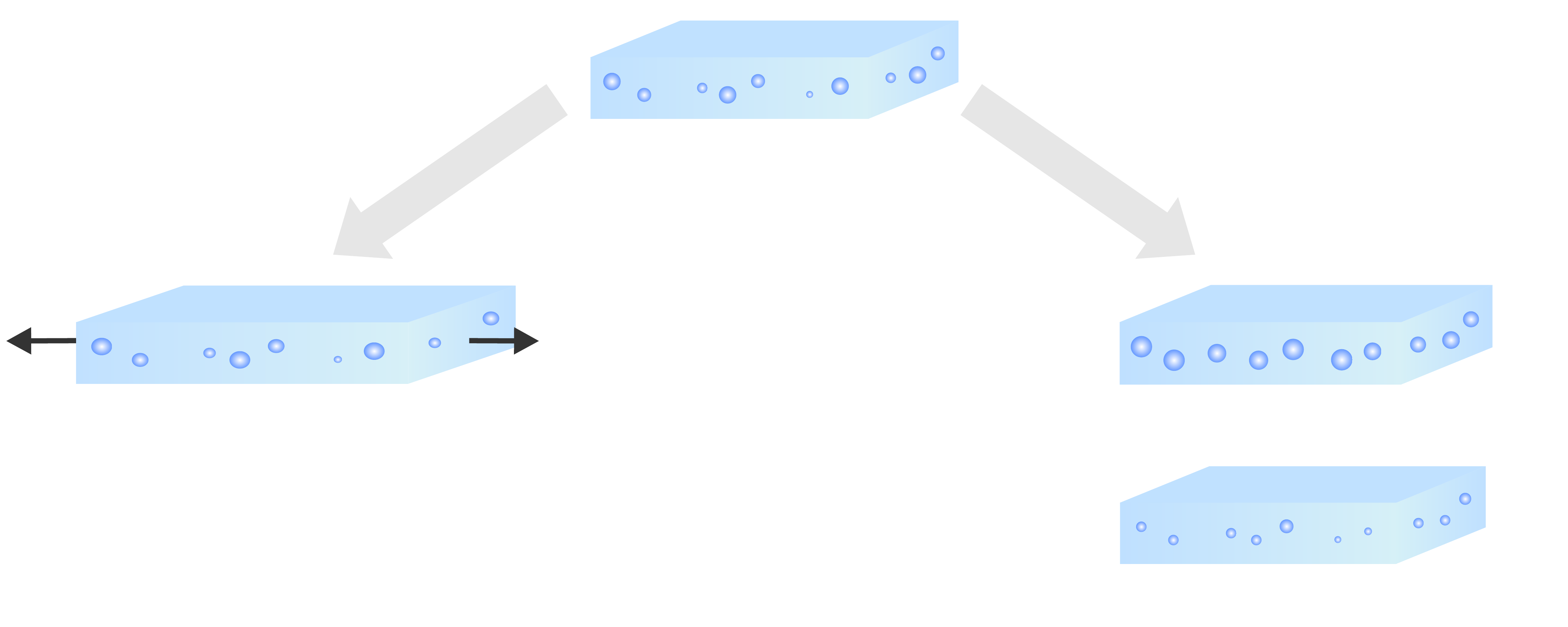}};
    \node[] at (-0.4,1.3) {$T=T^{mix}$};
    \node[] at (4.5,-2.6) {$T>T^{mix}$};
    \node[] at (4.5,-0.95) {$T<T^{mix}$};
    \node[] at (-4.5,-1) {$T=T^{mix}$};
    \node[] at (-7,-0.7) {$\epsilon_{xx}^{\infty}$};
    \node[] at (-2.3,-0.7) {$\epsilon_{xx}^{\infty}$};
\end{tikzpicture}
\caption{Sketch of the experimental set-up. Both deformation modes of uniaxial strain (left) and applied pressure (right) follow from the same starting structure.}
\label{fig:Sketch_Exp}
\end{figure}
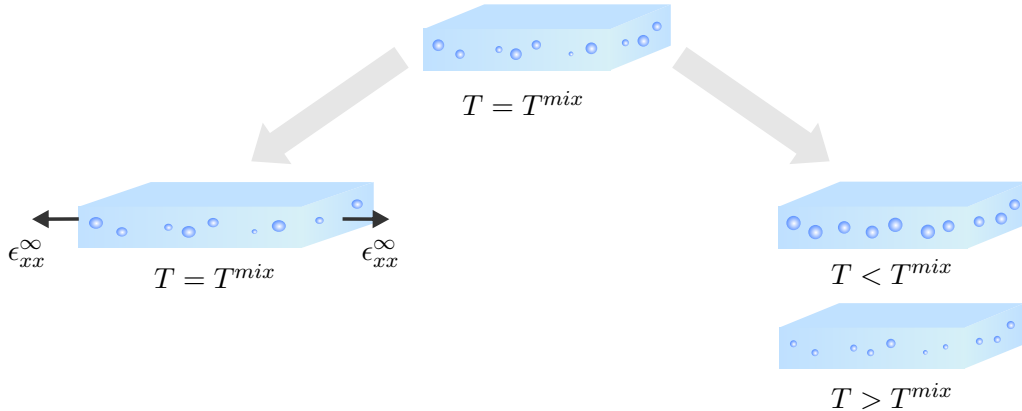

For the first test, 
the temperature is  fixed  and the sample is stretched \cite{Style:20152}, as shown on the left of Figure~\ref{fig:Sketch_Exp}.
Initial droplet radii are recorded and the  final cross-sectional shape of droplets are measured at a fixed uniaxial far-field strain. 
Droplets and matrix are assumed to be incompressible. 
The equilibrium shapes of the droplets are determined by a balance of elastic and interfacial stresses with the hydrostatic pressure, $p$, within the droplets.
The hydrostatic pressure $p$ may be decomposed as $p=p^0+p^{in}$, where $p^0$ is the Laplace pressure in the undeformed configuration and $p^{in}$ is an incremental  pressure increase, induced by deformation. 

For the second test, we simply shift the temperature. 
This changes the saturation concentration in the solid phase, which in turn induces a driving pressure $p^{in}$.
We record the subsequent change in droplet radii, as shown in the right of Figure 1.
When the temperature dependence of the saturation concentration is known, the driving pressure can be determined thermodynamically,
\cite{Style:2018,Rosowski:2020},
\begin{equation}
    p^{in} = n_Lk_BT\,\text{ln}\left(\frac{n^{sat}(T,p^{in})}{n^{sat}(T,0)}\right).
\label{eq:p}
\end{equation}
Here, $n_L$ is the number density of molecules in the condensed phase, and $k_B$ denotes Boltzmann's constant.  $n^{sat}(T,p^{in})$ denotes the equilibrium number density of molecules in the dilute phase, whereas $n^{sat}(T,0)$ denotes its equilibrium value in the absence of a mechanical load ($p^{in}=0$).

\section{Theoretical Background}

We seek to derive the deformation of an initially spherical closed elastic interface embedded in an elastic matrix under two different types of applied loading. 
The matrix is assumed to be incompressible. 
Matrix elasticity is modeled as a linear elastic isotropic solid exhibiting surface elasticity at the droplet/matrix interface. 
The outer matrix boundary is assumed to be far from the droplet/matrix interface, such that finite size effects of the elastic matrix may be neglected as well as the matrix deformation sufficiently far away from the droplet.
Droplet concentrations are furthermore taken to be in the dilute limit, such that interactions between droplets are negligible.

The equilibrium equations of a solid taking into account both bulk- as well as surface energies follows from the energy functional
\begin{equation}
    \mathcal{F}[\boldsymbol{\varphi}] = \int_{\Omega}W(\boldsymbol{\nabla}\boldsymbol{\varphi})\,dV + \int_{\partial\Omega^{\gamma}}\gamma(\boldsymbol{\nabla}\boldsymbol{\varphi})\,dS - \int_{\partial\Omega^t}\mathbf{t}_0\cdot\boldsymbol{\varphi}\,dS.
\end{equation}
Here, $\boldsymbol{\varphi}(\mathbf{X})$ is the deformation mapping with $\boldsymbol{\varphi}(\mathbf{X})\in\mathcal{U}=\{\boldsymbol{u}\in H^1(\Omega):\boldsymbol{\varphi}=\boldsymbol{\varphi}_0\,\text{on}\,\partial\Omega^u\}$, where $\partial\Omega^u$ is the boundary over which displacements are imposed. 
$W(\nabla\boldsymbol{\varphi})$ is the strain energy density of the bulk material, $\gamma(\nabla\boldsymbol{\varphi})$ is the strain energy density associated with the surface $\partial\Omega^{\gamma}$, and $\mathbf{t}_0$ is the applied traction along the traction boundary $\partial\Omega^t$. 
For a system to be in equilibrium, the principal of minimum potential energy requires a vanishing first variation of the energy functional $\delta\mathcal{F}[\boldsymbol{\varphi}]=0$ (for a detailed review on the derivation of governing equations within the framework of continuum mechanics, we refer to \cite{Huang:2013,Sharma:2020} and references therein). 
This results in the set of governing equations
\begin{subequations}
\begin{align}
    \text{div}(\boldsymbol{\sigma}) = 0 \quad &\text{in}\,\Omega, \label{eq:BVP_Bulk}\\
   \boldsymbol{\sigma}\mathbf{n} = \mathbf{t}_0 \quad &\text{on}\,\partial\Omega^t, \label{eq:BVP_Trac_Boundary} \\
   \boldsymbol{\sigma}\mathbf{n} - \text{div$_s$}(\boldsymbol{\sigma^s}) = \mathbf{t}_0 \quad &\text{on}\,\partial\Omega^{\gamma}, \quad \text{and} \label{eq:BVP_Boundary} \\
   \boldsymbol{u} = \boldsymbol{u}_0 \quad &\text{on}\,\partial\Omega^u \label{eq:BVP_Disp_Boundary}.
\end{align}
\label{eq:BVP}
\end{subequations}
Here, $\boldsymbol{\sigma}=\frac{\partial W}{\partial(\nabla\boldsymbol{\varphi})}$ and $\boldsymbol{\sigma^s}=\frac{\partial \gamma}{\partial(\nabla\boldsymbol{\varphi})}$ are the first Piola-Kirchhoff stress measures within the bulk- and surface material, respectively.
$\mathbf{n}$ is the outward unit normal to the surface, which is a function of surface displacements \cite{Vlahovska:2005}. 
As usual, displacements are defined as $\mathbf{u}=\boldsymbol{\varphi}(\mathbf{X})-\mathbf{X}$.
A salient feature of the proposed experiments is their applicability in the small deformation regime $|\boldsymbol{\nabla}_s\mathbf{u}|\sim\epsilon\ll 1$.
The matrix material is cured around pre-existing incompressible droplets. 
Therefore, the reference configuration is close to a critical stress-free configuration.
For an isotropic material, the linearized first Piola-Kirchhoff surface stress follows as \cite{Sharma:2019}
\begin{equation}
    \boldsymbol{\sigma}^s = \Upsilon^0\mathbf{I}^s + 2\mu^s\mathbf{E}^s + (\kappa^s-\mu^s)\text{tr}(\mathbf{E}^s)\mathbf{I}^s,
\label{eq:stress_strain_linear}
\end{equation}
where it is important to note that the residual surface stress $\Upsilon^0$ is of order $\mathcal{O}(\epsilon)$ \cite{Sharma:2020}.
Here, $\mathcal{O}(\epsilon^n)$ implies the asymptotic behavior $\mathcal{O}(\epsilon^n)/\epsilon^n\rightarrow C\neq 0$ as $\epsilon\rightarrow 0$, with $n\in\mathbb{R}$.  

\begin{figure}[ht]
\begin{tikzpicture}
    \node (exp_sym) at (0,0) {\includegraphics[width=0.85\textwidth]{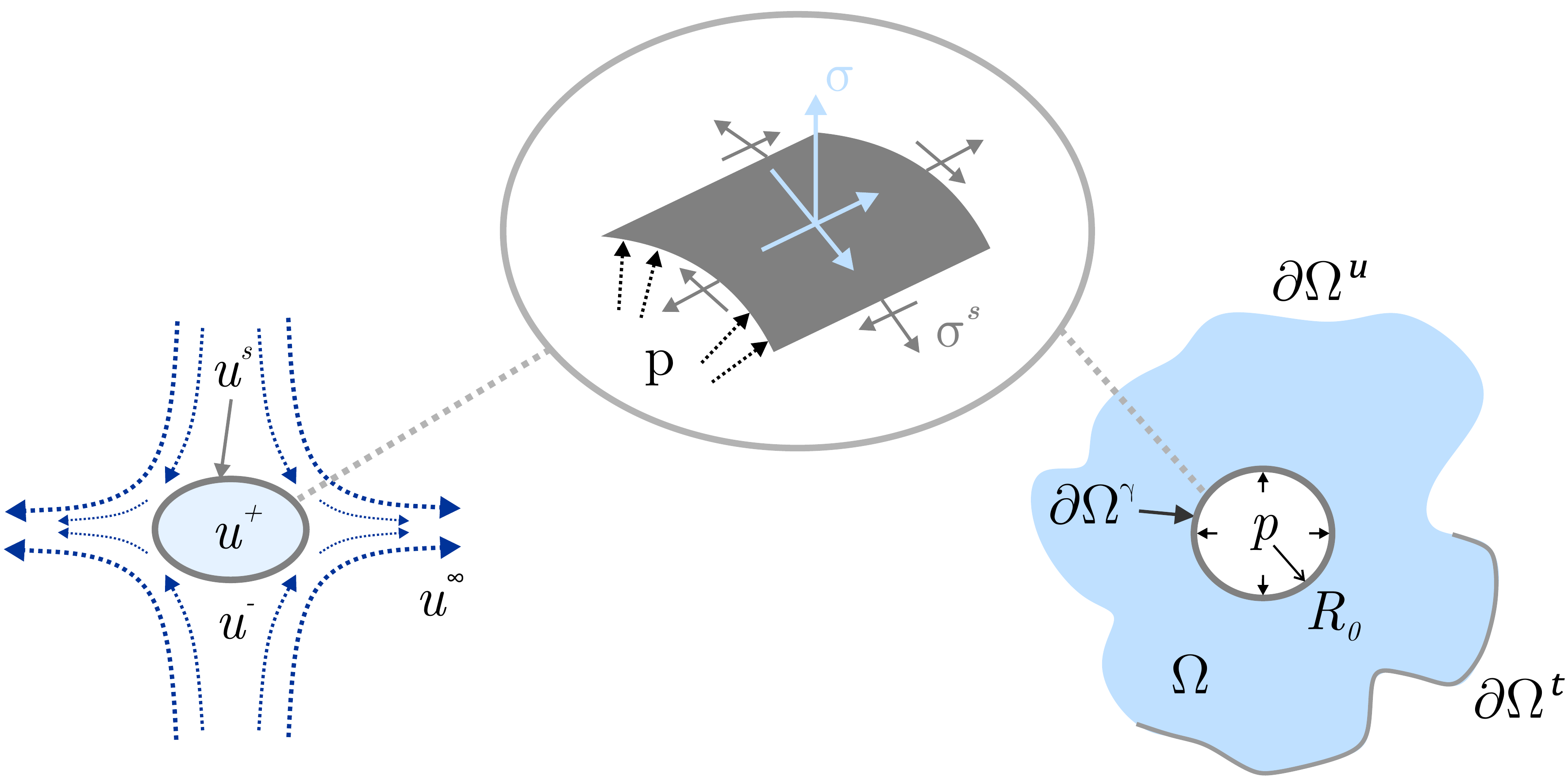}};
\end{tikzpicture}
\caption{Left: Elastic interface subjected to a displacement field $\mathbf{u}^{\infty}$ at infinity. Right: Infinite elastic matrix with an embedded cavity subjected to an inflation pressure $p$ at $r=R_0$. Both interfaces (gray) exhibit exhibit surface elasticity as highlighted in the inset.}
\label{fig:Schematic_BVP}
\end{figure}

\section{Shape of Droplets under Applied Stretch}
\label{sec:stretch}

We consider a static uniaxial strain applied at the far-field boundary of an elastic matrix with an embedded droplet exhibiting a hydrostatic state of stress. 
The droplet is composed of an incompressible liquid, hence the droplet's volume is preserved under the applied deformation.
The droplet/matrix interface is modeled as an initially spherical closed elastic interface.
The set of governing equations~\eqref{eq:BVP} reduces to
\begin{subequations}
\begin{align}
    \text{div}(\boldsymbol{\sigma}) = 0 \quad &\text{in}\,\Omega, \label{eq:BVP_Bulk_preserving}\\
   \boldsymbol{\sigma}\mathbf{n} = \mathbf{0} \quad &\text{on}\,\partial\Omega^t, \label{eq:BVP_Trac_Boundary_preserving} \\
   \boldsymbol{\sigma}\mathbf{n} - \text{div$_s$}(\boldsymbol{\sigma}^s) = \mathbf{0} \quad &\text{on}\,\partial\Omega^{\gamma}, \quad \text{and} \label{eq:BVP_Boundary_preserving} \\
   \mathbf{u} = \mathbf{u}^{\infty} \quad &\text{on}\,\partial\Omega^u \label{eq:BVP_Disp_Boundary_preserving}.
\end{align}
\label{eq:BVP_preserving}
\end{subequations}
Inspired by the dual nature of soft solids, we exploit the equivalence between linear elastic statics and Stokes flow of Newtonian fluids (see Appendix~\ref{sec:Equiv} for details). 
As illustrated in Figure~\ref{fig:Schematic_BVP}, we may thus adopt solutions for an elastic interface in a straining flow field \cite{Vlahovska:2016,Biesel:1981,Biesel:1985,Blawzdziewicz-Vlahovska-Loewenberg:2000} for the solution of the elastic problem.

\subsection{Interfacial shape equations}

The displacement field $\mathbf{u}^{\infty}$ at infinity induces disturbed displacement fields in the interface vicinity, which are denoted as $\mathbf{u}^{-}$ for the surrounding- and $\mathbf{u}^{+}$ for the encapsulated solid. 
In addition, $\mathbf{u^s}$ is the induced deformation of the elastic interface.
A triaxial affine strain follows from a displacement field $\mathbf{u}^{\infty}=(c_xx,c_yy,c_zz)$, which in spherical coordinates reads
\begin{align}
    \mathbf{u}^{\infty} &= r\left(c_z\cos(\theta)^2+c_x\cos(\varphi)^2\sin(\theta)^2+c_y\sin(\theta)^2\sin(\varphi)^2\right)\mathbf{g}_r \nonumber \\
    &+ r\left(-c_z\cos(\theta)\sin(\theta)+c_x\cos(\theta)\cos(\varphi)^2\sin(\theta)+c_y\cos(\theta)\sin(\theta)\sin(\varphi)^2\right)\mathbf{g}_{\theta} \nonumber \\
    &+ r\left(-c_x\cos(\varphi)\sin(\theta)\sin(\varphi)+c_y\cos(\varphi)\sin(\theta)\sin(\varphi)\right)\mathbf{g}_{\varphi}.
\label{eq:straining_flow}
\end{align}
Here, ($\mathbf{g}_r$, $\mathbf{g}_{\theta}$, $\mathbf{g}_{\varphi}$) denote the radial-, polar- and azimuthal base vectors, respectively. 

In the following, we exploit the spherical geometry of the problem and expand all variables in terms of spherical harmonics basis functions (see Appendix~\ref{Harmonics} for details). 
The displacement field of the interface is written as
\begin{align}
    \mathbf{u}^s &= f(\theta,\varphi)\mathbf{g}_r + \mathbf{u}^s_{||}, \nonumber \\
    f(\theta,\varphi) &= \sum_{j=1,j\neq 1}^{\infty}\sum_{m=-j}^jf_{jm}Y_{jm}(\theta,\varphi) \nonumber \\
    \mathbf{u}^s_{||}(\theta,\varphi) &= \sum_{j=1}^{\infty}\sum_{m=-j}^j(z_{jm}\mathbf{y}_{jm1}(\theta,\varphi) + u_{jm}\mathbf{y}_{jm0}(\theta,\varphi)),
\end{align}
where $\mathbf{u}^s_{||}$ is the displacement component tangential to the interface, $Y_{jm}$ are scalar spherical harmonics and $\mathbf{y}_{jmq}$ are vector spherical harmonics. 
For the displacement field satisfying Equation~\eqref{eq:final_solid} within the spherical harmonics basis, we adopt the fundamental solutions of the Stokes equations in a spherical geometry. 
We hence have
\begin{align}
    \mathbf{u}^- &= \sum_{jmq}c_{jmq}^-\hat{\mathbf{u}}_{jmq}^-(\mathbf{r}), \nonumber \\
    \mathbf{u}^+ &= \sum_{jmq}c_{jmq}^+\hat{\mathbf{u}}_{jmq}^+(\mathbf{r}) \quad \text{and}\nonumber \\
    \mathbf{u}^{\infty} &= \sum_{jmq}c_{jmq}^{\infty}\hat{\mathbf{u}}_{jmq}^+(\mathbf{r}).
\end{align}
Here, $\hat{\mathbf{u}}^{\pm}_{jmq}$ are vector solid spherical harmonics corresponding to the harmonics in the Lamb solution. 
Suffix notation with summation over repeated indices is implied. 
$q$ takes values $(0,1,2)$. 
$\hat{\mathbf{u}}^{\pm}_{jm2}$ is the radial component of the velocity field, while $\hat{\mathbf{u}}^{\pm}_{jm0}$ and $\hat{\mathbf{u}}^{\pm}_{jm1}$ are the tangential- and surface-solenoidal components, respectively.
Detailed expressions for flow fields $\hat{\mathbf{u}}^{\pm}_{jmq}$ and resultant tractions in terms of spherical harmonics are listed in Appendix~\ref{sec:fundamental_solutions}.
The coefficients $c^{\infty}_{jmq}$ for a generic straining flow as given in Equation~\eqref{eq:straining_flow} are 
\begin{align}
    c^{\infty}_{200} &= \sqrt{\frac{6\pi}{5}}c_z, \, c^{\infty}_{220} = \sqrt{\frac{\pi}{5}}(c_x-c_y) \nonumber \\
    c^{\infty}_{202} &= 2\sqrt{\frac{\pi}{5}}c_z, \, c^{\infty}_{222} = \sqrt{\frac{2\pi}{15}}(c_x-c_y).
\end{align}
At linear perturbation order \cite{Vlahovska:2016}, interfacial elastic stresses following from Equation~\eqref{eq:stress_strain_linear} may be written as
\begin{align}
    \tau^s_{jm0} &= -2C_{\kappa^s}^{-1}\sqrt{j(j+1)}f_{jm} + \left((C_{\mu^s}^{-1}+C_{\Upsilon^0}^{-1})(j-1)(j+2) + C_{\kappa^s}^{-1}j(j+1)\right)u_{jm}, \nonumber \\
    \tau^s_{jm1} &= -(C_{\mu^s}^{-1}+C_{\Upsilon^0}^{-1})(j-1)(j+2)z_{jm}, \quad \text{and} \nonumber \\
    \tau^s_{jm2} &= C_{\Upsilon^0}^{-1}(j(j+1)-2)f_{jm} + 2C_{\kappa^s}^{-1}\left(\sqrt{j(j+1)}u_{jm}-2f_{jm}\right),
\end{align}
with non-dimensionalized material constants
\begin{equation}
    C_{\kappa^s}=\frac{\mu R_0}{\kappa^s}, \, C_{\mu^s}=\frac{\mu R_0}{\mu^s}, \, C_{\Upsilon^0}=\frac{\mu R_0}{\Upsilon^0}.
\end{equation}
Here, $R_0$ is the initial radius of the elastic interface.
Finally, shape parameters $(u_{jm},z_{jm},f_{jm})$ are determined from the continuity of normal tractions at the interface as specified in Equation~\eqref{eq:BVP_Boundary_preserving}. 
For comparison with experiment, normalized displacements parallel and perpendicular to the applied strain are needed. Normalized displacements follow as
\begin{align}
    \frac{\mathbf{u}^s_{(\theta=0,\varphi=0)}}{R_0} &= \frac{5ER_0(2ER_0+9\kappa^s+3\mu^s)c_z}{3\left(2E^2R_0^2+12\mu^s\kappa^s+6\Upsilon^0(3\kappa^s+2\mu^s)+ER_0(5\Upsilon^0+8\kappa^s+6\mu^s)\right)}\mathbf{g}_r, \nonumber \\
   \frac{\mathbf{u}^s_{(\theta=\pi/2,\varphi=0)}}{R_0} &= \frac{5ER_0(2ER_0+9\kappa^s+3\mu^s)(c_x-c_y-c_z)}{6\left(2E^2R_0^2+12\mu^s\kappa^s+6\Upsilon^0(3\kappa^s+2\mu^s)+ER_0(5\Upsilon^0+8\kappa^s+6\mu^s)\right)}\mathbf{g}_r.
 \label{eq:governing_asym}
\end{align}
Appendix~\ref{sec:disp_appendix} gives the full form of the interfacial displacement field. To linear order, displacements ($\mathbf{u}^s_{(\theta=0,\varphi=0)}$,$\mathbf{u}^s_{(\theta=\pi/2,\varphi=0)}$) suffice to determine the resultant ellipsoidal interface shape, where $(l=2(R_0+\mathbf{u}^s_{(\theta=0,\varphi=0)}),w=2(R_0+\mathbf{u}^s_{(\theta=\pi/2,\varphi=0)}))$ are the length- and width of the droplet in the $(x,z)$-plane, respectively.
We can furthermore define the ratio of microscopic to macroscopic strains, which is invariant under arbitrary biaxial loading conditions, as
\begin{equation}
    \frac{l-w}{R_0(c_z-c_x)} = \frac{10ER_0(2ER_0+9\kappa^s+3\mu^s)}{3\left(2E^2R_0^2+12\mu^s\kappa^s+6\Upsilon^0(3\kappa^s+2\mu^s)+ER_0(5\Upsilon^0+8\kappa^s+6\mu^s)\right)}.
\label{eq:drop_strain}
\end{equation}

\begin{figure}[ht]
 \includegraphics[width=0.8\textwidth]{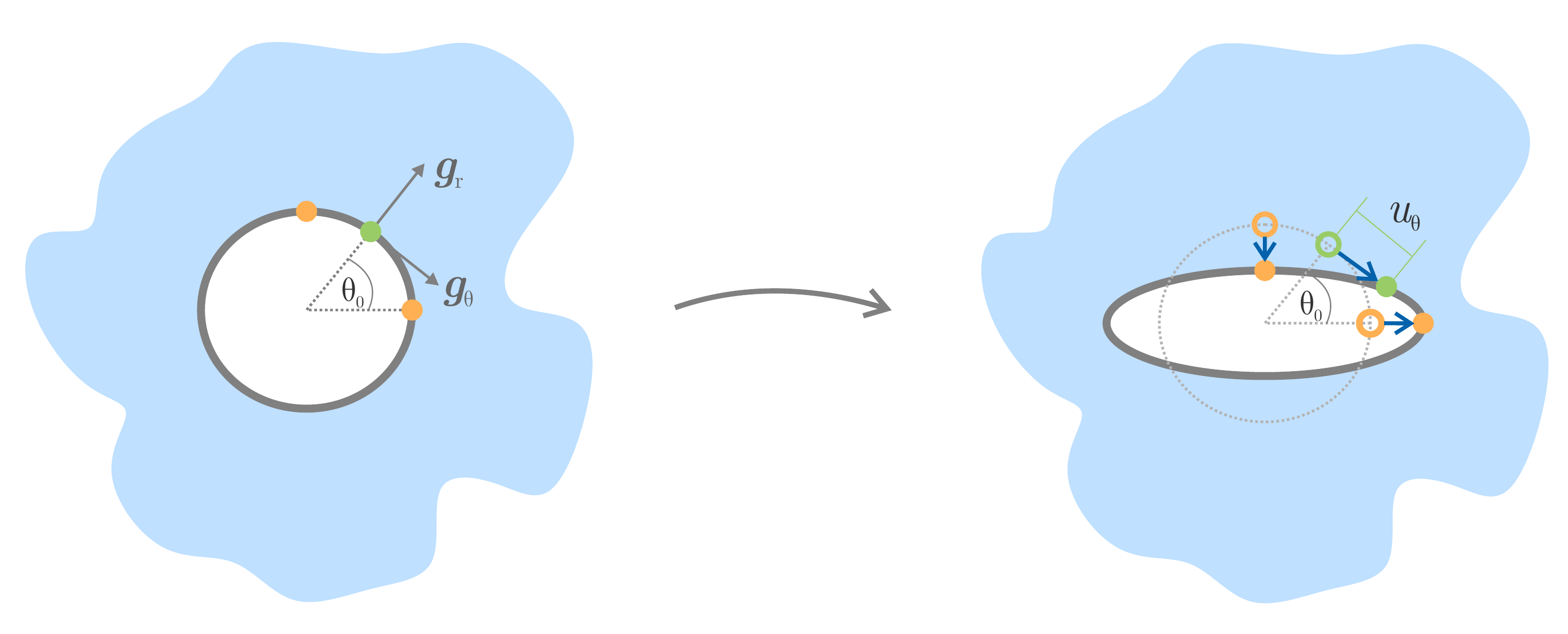}
\caption{Geometrical construction to determine hoop displacements at distinct surface points located at $\theta=\theta_0$ under axisymmetric loading $c_x=c_y$.}
\label{fig:Construction}
\end{figure}

The above measures only depend on the radial component of the interfacial displacement field.
To add more information to the subsequent fitting, it is beneficial to also consider a measure that depends on interfacial displacement within the hoop direction.
In experiments, this usually requires to track points at the droplet surface. 
We circumvent this by noting that there is a distinct angle at which radial displacements vanish. 
Here, the resultant deformation can be purely attributed to hoop displacements (shown as green points in Figure~\ref{fig:Construction}).
Using axisymmetric loading conditions $c_x=c_y$, Equation~\eqref{eq:disp_r_appendix} gives 
\begin{equation}
    \theta_0\coloneqq\theta|_{u^s_r=0} = \frac{\text{arccos}(-1/3)}{2}.
\label{eq:theta}
\end{equation}
Points at the initial droplet surface at an angle $\theta_0$ solely deform within the hoop direction, thus providing a unique mapping to deformed points at the ellipsoidal surface.
The hoop displacement, $u^s_{\theta}$ at $\theta_0$ can be determined trigonometrically from the values of $(R_0,l,w)$, as shown in Figure~\ref{fig:Construction}.

This is related to the material properties through Equation ~\eqref{eq:disp_r_appendix}, 
\begin{equation}
    \frac{1}{R_0c_z}u^s_{\theta}|_{\theta=\theta_0} = -\frac{5ER_0(2ER_0+3\Upsilon^0+6\kappa^s)\cos{\theta_0}\sin{\theta_0}}{2\left(2E^2R_0^2+12\mu^s\kappa^s+6\Upsilon^0(3\kappa^s+2\mu^s)+ER_0(5\Upsilon^0+8\kappa^s+6\mu^s)\right)}.
\label{eq:hoop_disp}
\end{equation}
Quantities given in Equations~\eqref{eq:drop_strain} and~\eqref{eq:hoop_disp} are used in the fitting procedure outlined in Sections~\ref{sec:indi_fit} and~\ref{sec:Sensitivities}.

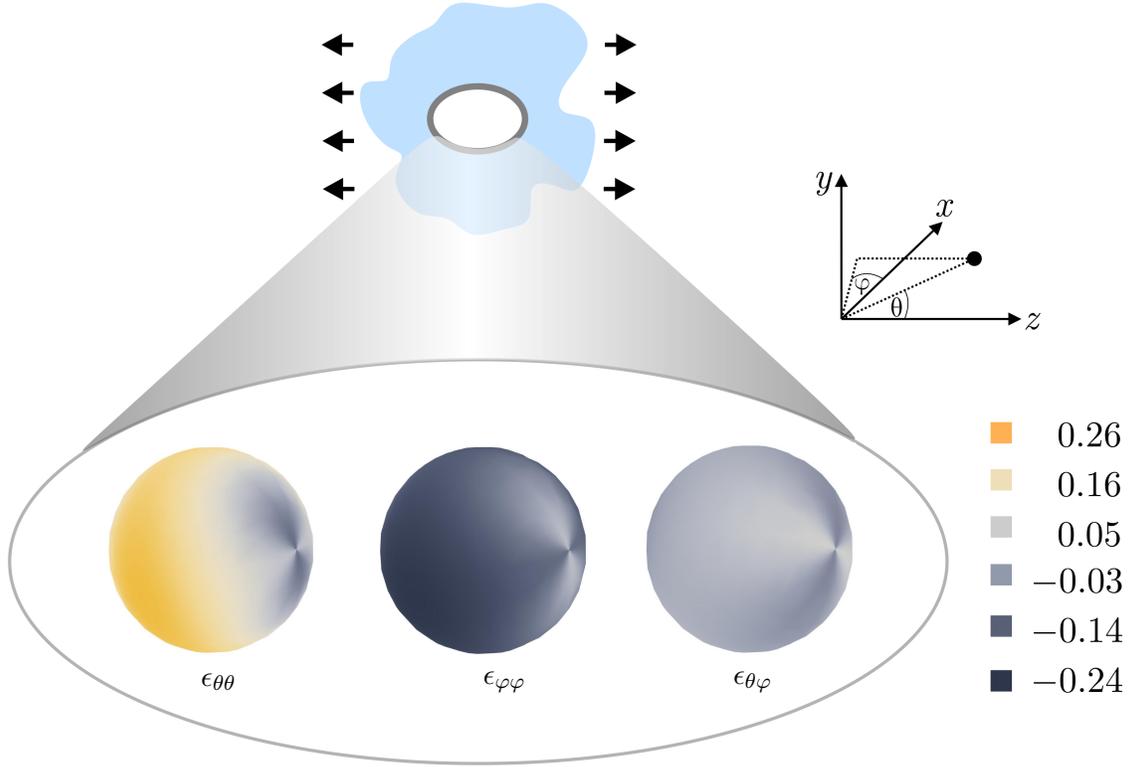
\begin{figure}[ht]
\begin{tikzpicture}
    \node (strains) at (0,0) {\includegraphics[width=0.9\textwidth]{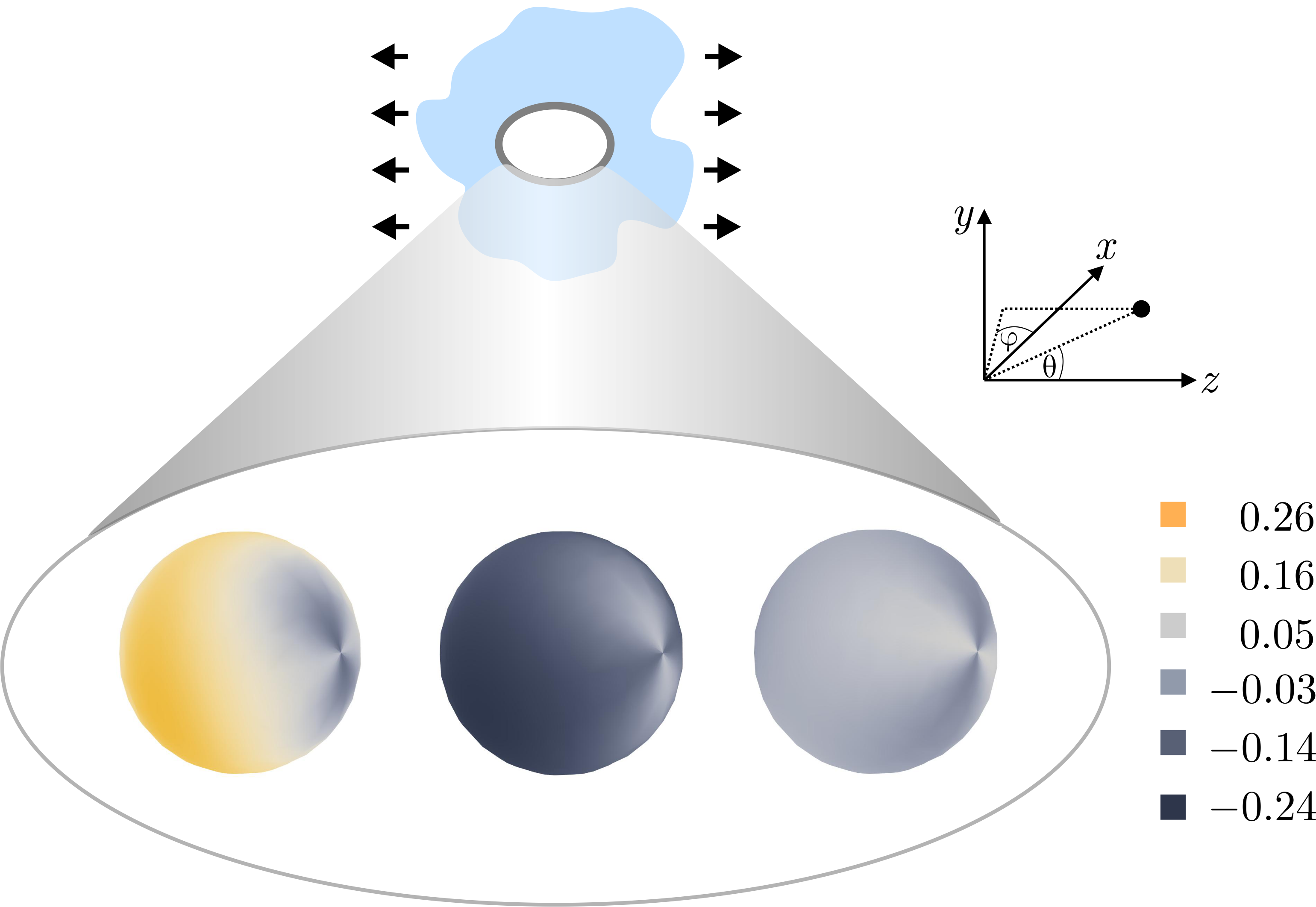}};
    \node[] at (-4.6,-4) {$\epsilon_{\theta\theta}$};
    \node[] at (-0.8,-4) {$\epsilon_{\varphi\varphi}$};
    \node[] at (2.5,-4) {$\epsilon_{\theta\varphi}$};
\end{tikzpicture}
\caption{Strain profiles at the interface of an incompressible droplet for an applied displacement field $\mathbf{u}^{\infty}=(-\frac{3}{4}x,-\frac{1}{4}y,z)$ at the far-field boundary of the elastic matrix. Material parameters are $E=1000$\,kPa, $\Upsilon^0=0.01$\,N/m, $\mu^s=0.01$\,N/m, $\kappa^s=0.02$\,N/m. 
}
\label{fig:Strains}
\end{figure}

Figure~\ref{fig:Strains} illustrates resultant strain profiles at the interface. 
To feature a fully generic deformation for which all components of surface strain vary, we choose an applied displacement of the form $\mathbf{u}^{\infty}=(-\frac{3}{4}x,-\frac{1}{4}y,z)$. 
In addition to strains $\epsilon_{\theta\theta}$ and $\epsilon_{\varphi\varphi}$ within the hoop directions, the compressive asymmetry in $\mathbf{u}^{\infty}$ also induces shear strains $\epsilon_{\theta\varphi}$. 
In the plane perpendicular to the applied loading, highest tensile strains are attained within the polar hoop direction $\epsilon_{\theta\theta}$, whereas the azimuthal hoop direction $\epsilon_{\varphi\varphi}$ exhibits highest compressive strains. 
Induced shear strains are of comparably low magnitude.

\subsection{Fitting droplet shapes under applied stretch}
\label{sec:indi_fit}

\begin{figure}[ht]
\begin{tikzpicture}
    \node (fit) at (0,0) {\includegraphics[width=0.9\textwidth]{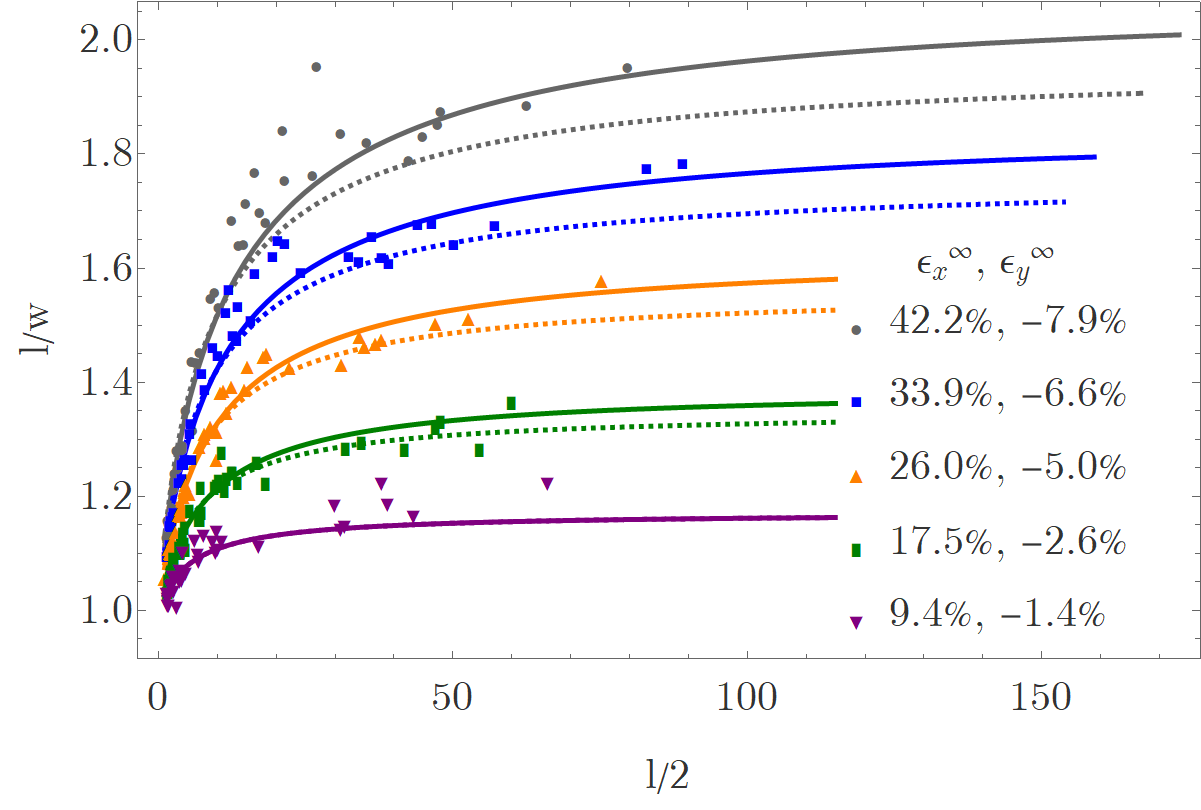}};
    \node (sketch) at (4.8,3.5) {\includegraphics[width=0.3\textwidth]{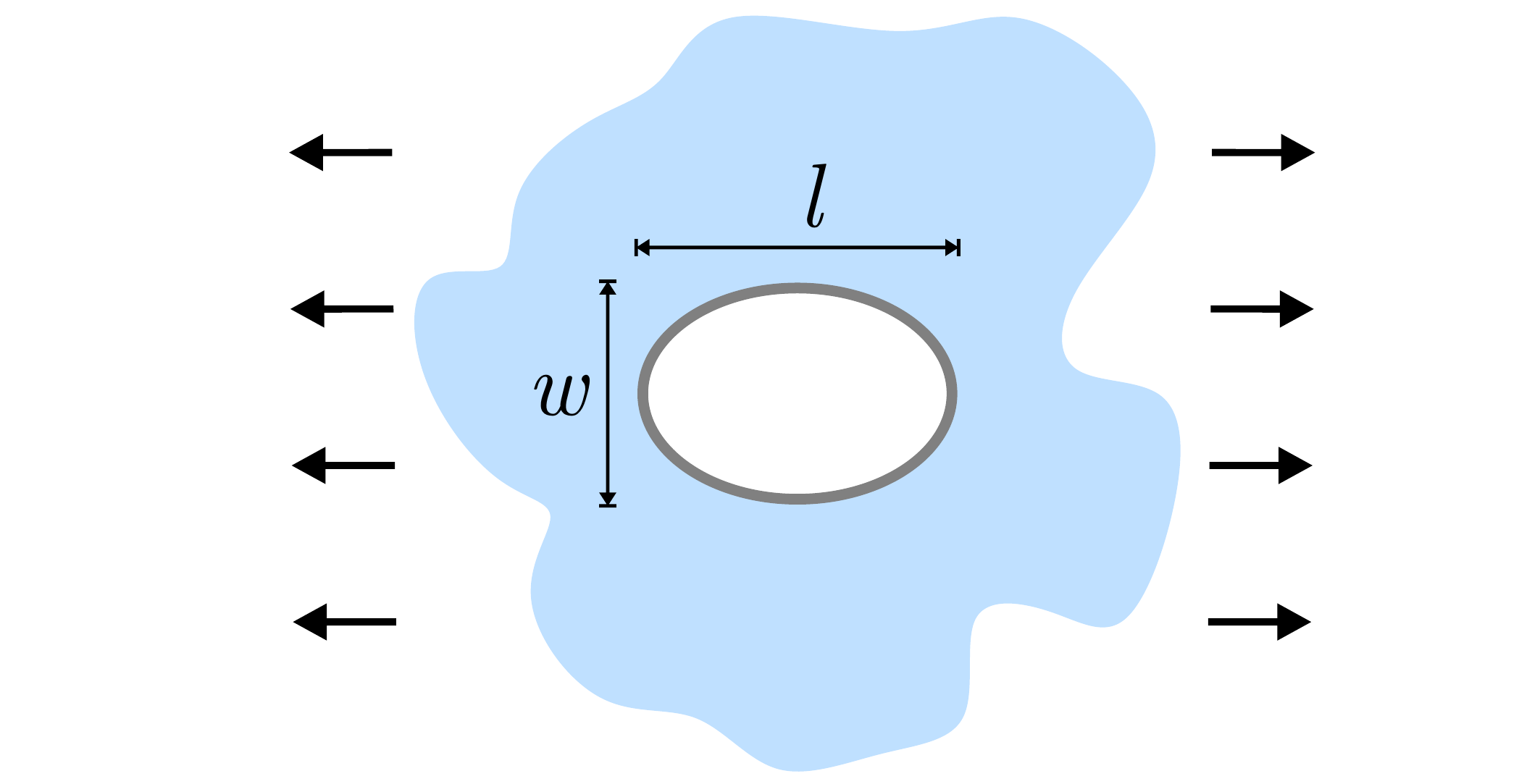}};
\end{tikzpicture}
\caption{Sample fitting of experimental data given in \cite{Style:20152} to Equation~\eqref{eq:governing_asym}. Dashed lines illustrate predictions for vanishing surface elasticity $(\Upsilon^0,0,0)$, whereas solid lines depict the behavior incorporating surface elastic effects $(\Upsilon^0,\mu^s,\kappa^s)$.}
\label{fig:Fitting}
\end{figure}

The effect of surface elasticity on the resulting deformation field may be highlighted in comparison to available experimental data \cite{Style:20152}. 
Under the assumption of constant surface energy, a best-fit value of residual surface stress $\Upsilon^0=3.6$\,mN/m is obtained (see dashed lines in Figure~\ref{fig:Fitting}). 
In contrast, taking the complete set of surface material parameters into account, best-fit values are ($\Upsilon^0=3.9\pm 1.7$\,mN/m, $\mu^s=2.0\pm 2.6$\,mN/m, $\kappa^s=510\pm 280$\,mN/m). 
In both cases, a value of $E=1.7\,\mathrm{kPa}$ is used for the material's bulk Young's modulus.
While introducing surface elastic effects does a better job fitting the data (see solid lines in Figure~\ref{fig:Fitting}), the uncertainty in these parameters is too large to be useful.

\begin{figure}[ht]
\begin{tikzpicture}
    \node (flow) at (-4,5.5) {\includegraphics[width=0.45\textwidth]{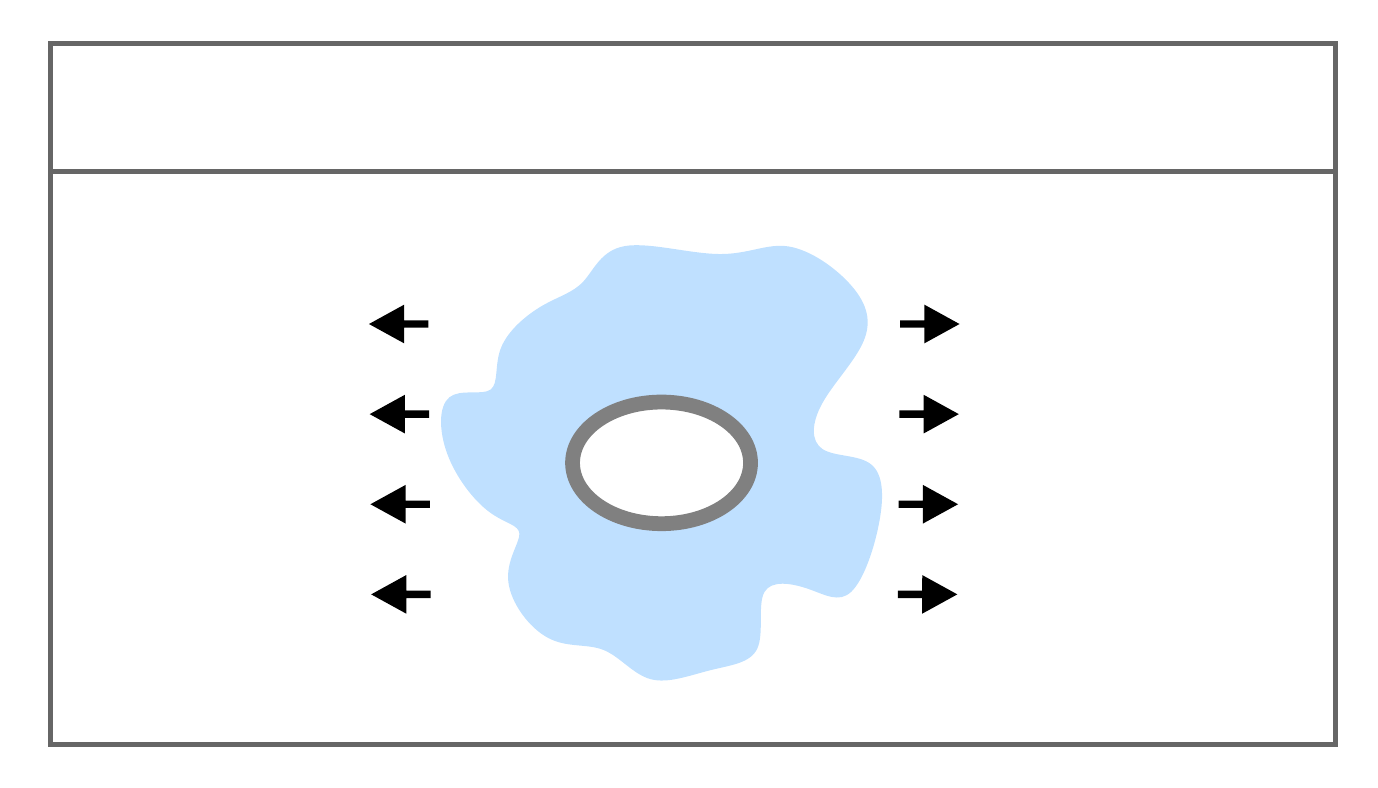}};
    \node[] at (-4,7) {$P(\Upsilon^0,\mu^s,\kappa^s|D^{strain}_1\cup D^{strain}_2)$};
    \node () at (4,5) {\includegraphics[width=0.45\textwidth]{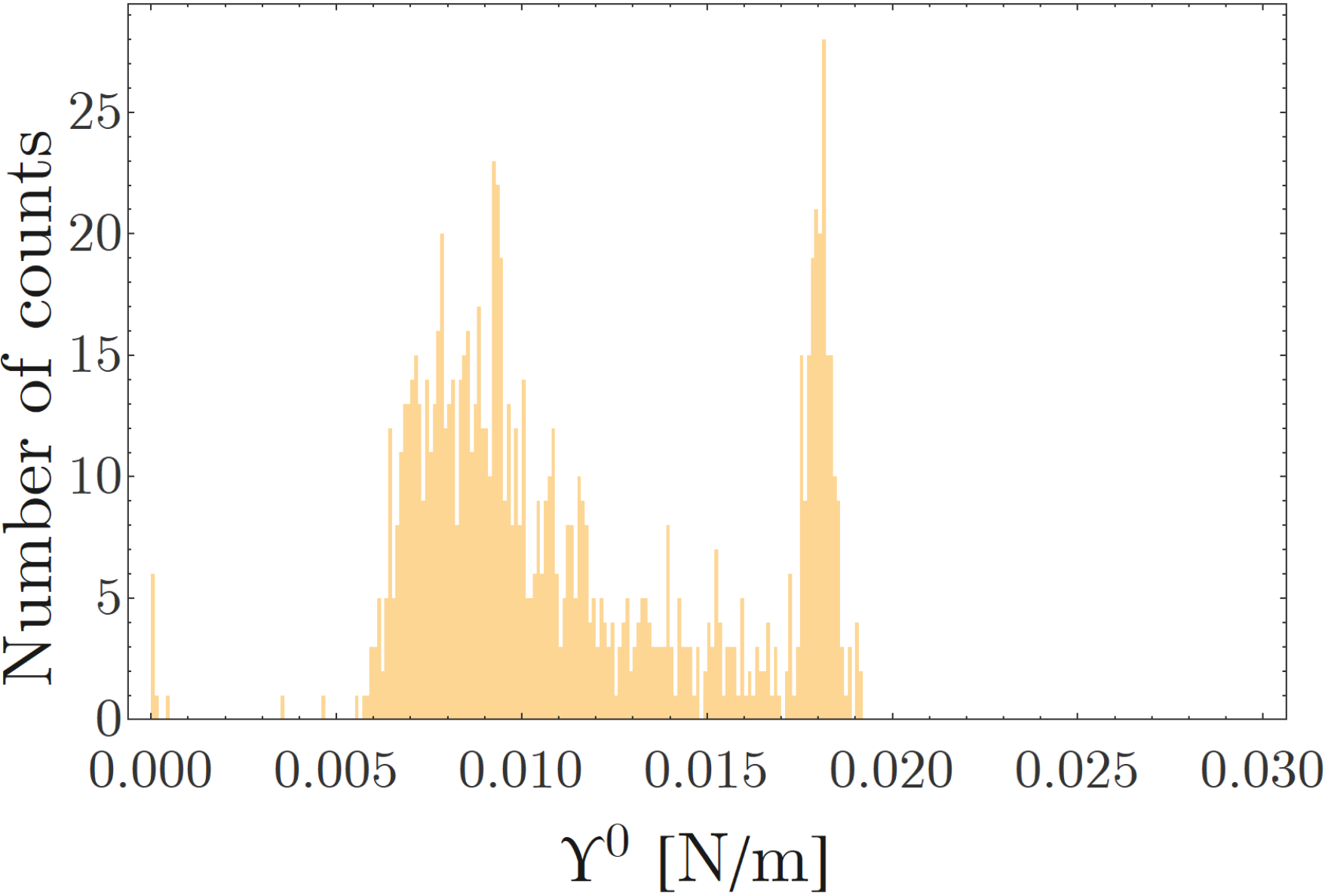}};
    \node () at (-4,-0.5) {\includegraphics[width=0.45\textwidth]{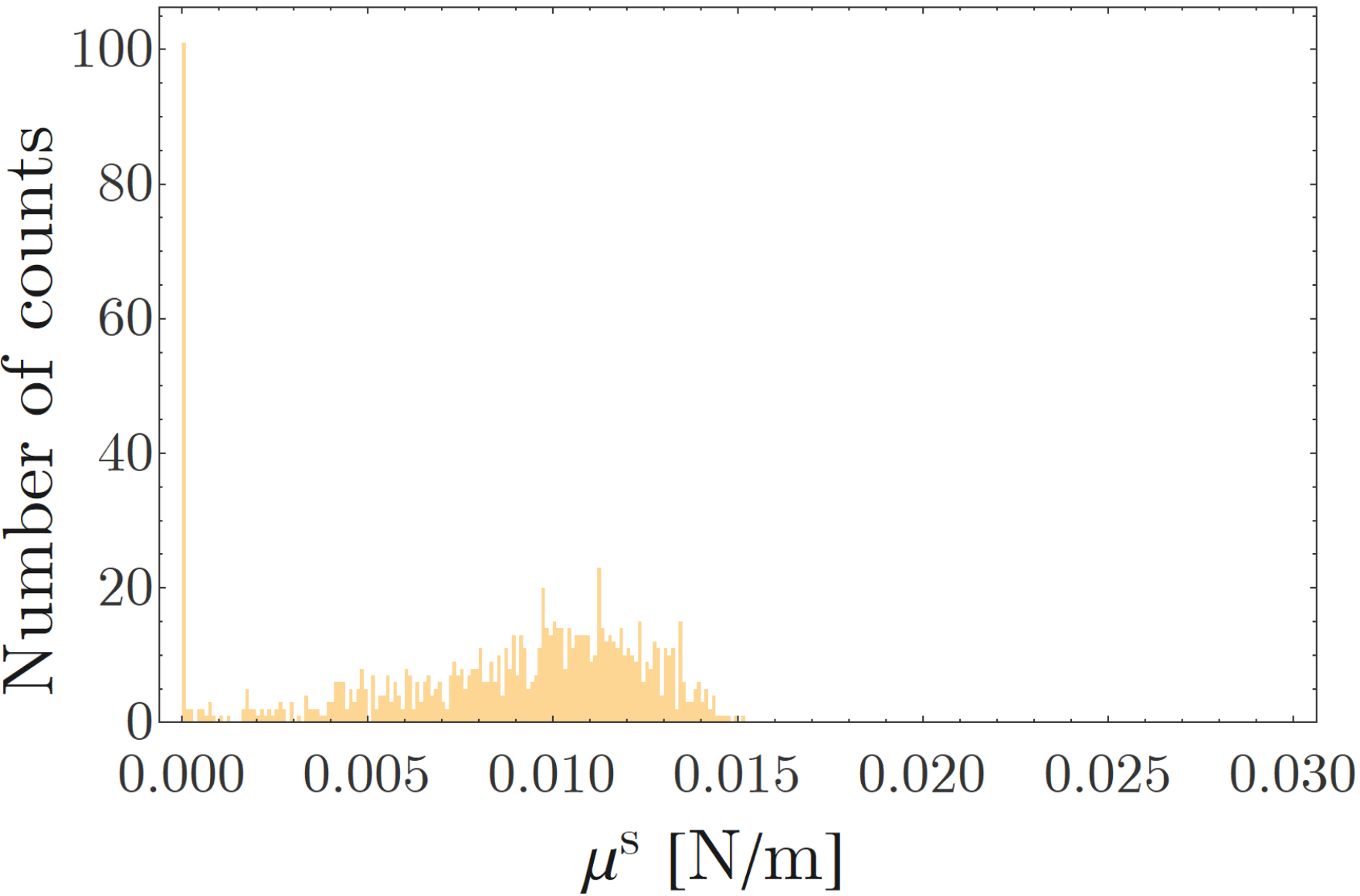}};
    \node () at (4,-0.5) {\includegraphics[width=0.45\textwidth]{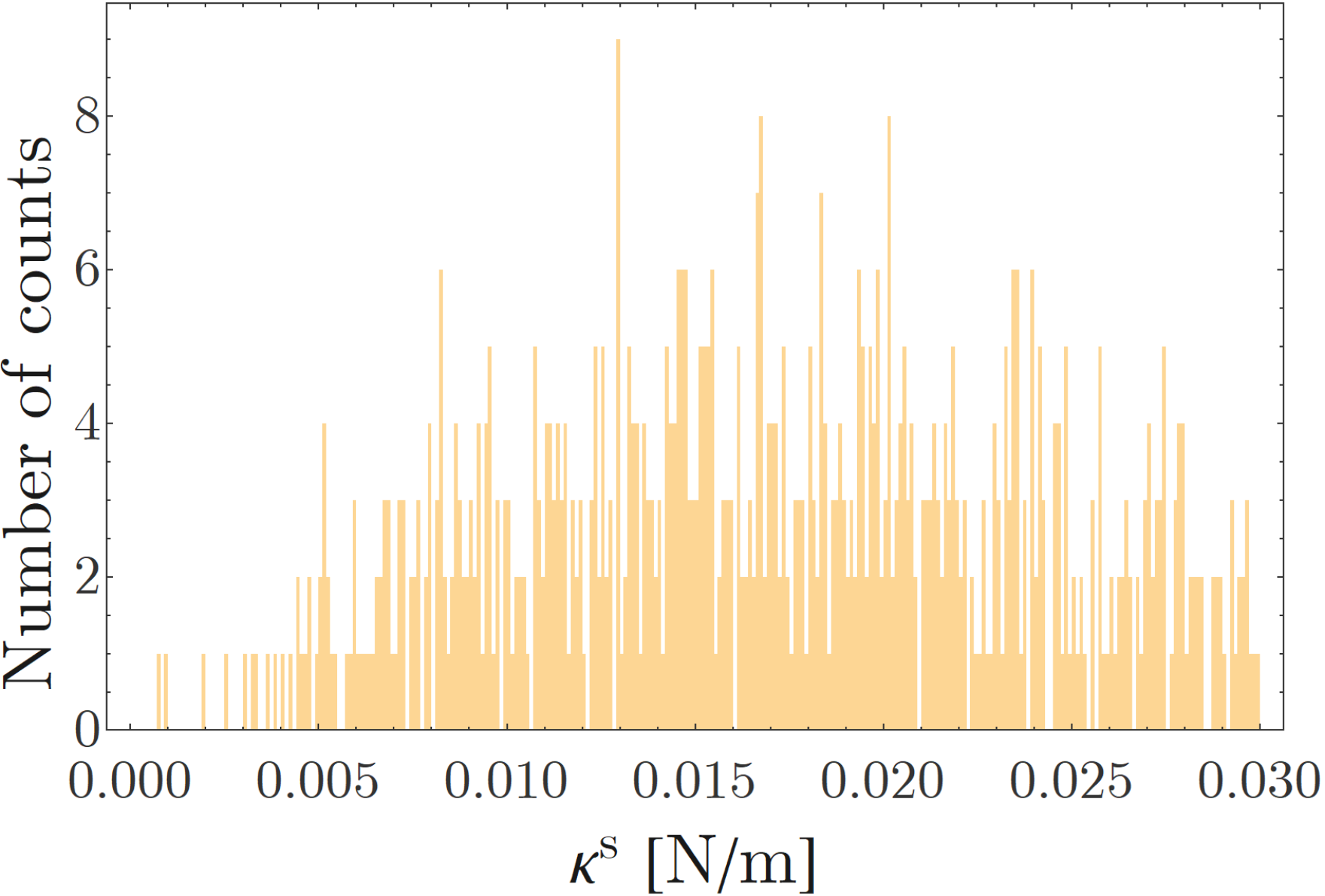}};
\end{tikzpicture}
\caption{Joint minimization of the volume preserving data set $D^{strain}_1\cup D^{strain}_2$ to Equations~\eqref{eq:drop_strain} and~\eqref{eq:hoop_disp} and resultant probability distribution $P(\Upsilon^0,\mu^s,\kappa^s|D^{strain}_u\cup D^{strain}_2)$ for test values $\Upsilon^0=0.01\,\text{N/m}$, $\mu^s=0.01\,\text{N/m}$, $\kappa^s=0.02\,\text{N/m}$.}
\label{fig:Fitting_Asym}
\end{figure}

To quantify the reliability of fitting parameters, we proceed with the generation of artificial data sets. 
We insert test parameters into the analytical solution given in Equations~\ref{eq:drop_strain} and~\ref{eq:hoop_disp}. 
The addition of gaussian noise results in artificial data sets $D^{strain}_1$ and $D^{strain}_2$, respectively.
Here, our assumption for the experimental error is based on available experimental data illustrated in Figure~\ref{fig:Fitting}, with relative standard deviation $\sim 12\%$.
Based on the calculated standard deviation, a pool of data sets $D^{strain}_1\cup D^{strain}_2$ and corresponding solutions to the nonlinear optimization problem is generated. 
As highlighted in Figure~\ref{fig:Fitting_Asym}, the probability distributions of resultant best-fit surface parameters $(\Upsilon^0,\mu^s,\kappa^s)$ show broad spectra and do not recover test parameters ($\Upsilon^0=0.01\,\text{N/m}$, $\mu^s=0.01\,\text{N/m}$, $\kappa^s=0.02\,\text{N/m}$). 
A reliable estimation of the complete set of surface material parameters hence calls for an additional mode of deformation to be taken into account.
Ideally, the additional deformation mode only depends on a subset of surface parameters $(\Upsilon^0,\mu^s,\kappa^s)$, which is achieved by considering a hydrostatic inflation as outlined in the following Section.

\section{Growth and shrinkage in an elastic matrix}
\label{sec:growth}

The growth of a condensing droplet under an applied pressure is modeled as a hydrostatic inflation of a spherical closed elastic interface embedded in an elastic matrix (see sketch in Figure~\ref{fig:Schematic_BVP}). 
In contrast to the uniaxial deformation, this hence constitutes a volume distorting deformation mode.
The set of governing equations~\eqref{eq:BVP} specifies to
\begin{subequations}
\begin{align}
    \text{div}(\boldsymbol{\sigma}) = 0 \quad &\text{in}\,\Omega, \label{eq:BVP_Bulk_distorting}\\
   \boldsymbol{\sigma}\mathbf{n} = \mathbf{0} \quad &\text{on}\,\partial\Omega^t\setminus\partial\Omega^{\gamma}, \label{eq:BVP_Trac_Boundary_distorting} \\
   \boldsymbol{\sigma}\mathbf{n} - \text{div$_s$}(\boldsymbol{\sigma^s}) = p\mathbf{n} \quad &\text{on}\,\partial\Omega^{\gamma}, \quad \text{and} \label{eq:BVP_Boundary_distorting} \\
   \mathbf{u} = \mathbf{0} \quad &\text{on}\,\partial\Omega^u \label{eq:BVP_Disp_Boundary_distorting}.
\end{align}
\label{eq:BVP_distorting}
\end{subequations}

Based on the spherical symmetry of the deformation, the unknown displacement field simplifies to $\mathbf{u}=u_r(r)\,\mathbf{g}_r$, where $\mathbf{g}_r$ is the radial base vector in spherical coordinates. Governing equation~\eqref{eq:BVP_Bulk_distorting} in combination with the constitutive choice of linear elastic and isotropic material behavior hence gives
\begin{equation}
    r^2\frac{\partial^2u_r}{\partial r^2} + 2r\frac{\partial u_r}{\partial r}-2u_r = 0 \quad \text{in}\,\Omega,
\label{eq:ODE_Bulk}
\end{equation}
which is similar to the Rayleigh-Plesset equation modeling pressure pulses within bubbles \cite{Strutt:1917,Plesset:1949}.
The general solution to Equation~\eqref{eq:ODE_Bulk} in terms of unknown constants $(C_1,C_2)$ is
\begin{equation}
    u_r(r) = C_1r + C_2r^{-2}, 
\end{equation}
where we require $C_1=0$ for a vanishing displacement field at infinity. 
It remains to determine $C_2$ using Equation~\eqref{eq:BVP_Boundary_distorting}. With
\begin{align}
    \mathbf{g}_r\cdot\text{div}_s(\boldsymbol{\sigma}^s)&=-\boldsymbol{\sigma}^s\cdot\boldsymbol{\nabla}_s\,\mathbf{g}_r \nonumber \\
    &= -\boldsymbol{\sigma}^s\cdot\left(\frac{1}{r}\left(\mathbf{g}_{\theta}\otimes\mathbf{g}_{\theta}+\mathbf{g}_{\varphi}\otimes\mathbf{g}_{\varphi}\right)\right),
\end{align}
where $\otimes$ denotes an outer product, Equation~\eqref{eq:BVP_Boundary_distorting} reduces to
\begin{equation}
    \sigma_{rr} = \frac{2\sigma^s_{\theta\theta}}{R_0} - p.
\end{equation}
We thus find
\begin{equation}
    C_2 = \frac{3R_0^3(pR_0-2\Upsilon^0)}{4(ER_0+3\kappa^s)},
\end{equation}
from which
\begin{equation}
    u_r = R-R_0 = \frac{3R_0(pR_0-2\Upsilon^0)}{4(ER_0+3\kappa^s)}
\label{eq:governing_sym_dim}
\end{equation}
determines the radial displacement at the inflation boundary $r=R_0$. 
Since the Laplace pressure has to be recovered in the limit of vanishing driving pressure $p^{in}$, we substitute $p=p^{in}+2\Upsilon^0/R_0$ and define the normalized radial displacement as
\begin{equation}
    \hat{u}_r = \frac{u_r}{R_0}\frac{E}{p^{in}} = \frac{3ER_0}{4(ER_0+3\kappa^s)}.
\label{eq:governing_sym}
\end{equation}

\section{Coupling mechanics to solubility}
\label{sec:growth_appendix}

In this Section, we  connect  the mechanics of deformations derived in Sections~\ref{sec:stretch} and~\ref{sec:growth} to the dissolution of droplets governed by Equation~\ref{eq:p}.  This lays the groundwork for assessing the 
  experimental viability of the tests for surface elasticity proposed in Section \ref{sec:Experiments}. 
  In particular, we focus on stretch-induced dissolution during uniaxial stretch (Section~\ref{sec:stretch}), and the sensitivity of droplet radius to changes of temperature  (Section~\ref{sec:growth}).

In both cases, the average number density of molecules,  
\begin{align}
	n^{av} &= \phi  n_L + (1-\phi) n^{sat}(T,p^{in}) ,
\label{eq:number_densities}
\end{align}
is constant throughout deformation/ temperature shifts. We thus have
\begin{align}
    &\frac{\partial n^{av}}{\partial n^{sat}(T,p^{in})}\frac{\partial n^{sat}(T,p^{in})}{\partial T} dT + \frac{\partial n^{av}}{\partial n^{sat}(T,p^{in})}\frac{\partial n^{sat}(T,p^{in})}{\partial p^{in}} dp^{in} 
    + \frac{\partial n^{av}}{\partial\phi}\frac{\partial\phi}{\partial R}dR = 0,
\label{eq:total_diff}
\end{align}
where $\phi=\mathcal{N}\cdot\frac{4}{3}\pi R^3$ is the volume fraction of the condensed phase and $\mathcal{N}$ the number density of droplets.

\subsection{Sensitivity of temperature}

In Section \ref{sec:Experiments}, we proposed to drive droplet growth through a temperature-induced shift in solubility.
The viability of this approach depends on 
a careful choice of droplet composition.  An appropriate system will have a temperature dependence of solubility that enables ready access to small but resolvable deformations.
Equation~\eqref{eq:governing_sym_dim} provides an expression for the mechanical pressure inducing droplet growth. In the initial configuration, we have $p^{in}=0$, $R=R_0$, and $n^{sat}(T,p^{in})=n^{sat}(T,0)$, such that  
\begin{equation}
    dp^{in} = \frac{4(ER+3\kappa^s)}{3R^2}dR.
\end{equation}
Evaluating the individual terms in Equation~\eqref{eq:total_diff} therefore gives
\begin{equation}
    \frac{1}{R}\frac{dR}{dT} = - ~ \frac{1-\phi}{3\phi(n_L-n^{sat}(T,0))+\frac{4(1-\phi)n^{sat}(T,0)(ER+3\kappa^s)}{3~ n_Lk_BT~ R}} ~ \frac{\partial n^{sat}(T,0)}{\partial T}.
\label{eq:dR_dT}
\end{equation}
As we can see from this equation, changes in droplet size are more sensitive to temperature shifts for small volume fractions of the condensed phase and large differential solubilities. In addition, the right hand side of Equation~\eqref{eq:dR_dT} is independent of droplet size for vanishing surface elasticity $\kappa^s=0$. 

\subsection{Stretch-induced dissolution}

Section~\ref{sec:stretch} considers a droplet in an elastic matrix under an applied far-field stretch. 
This deformation increases the pressure within the droplets, leading to a driving pressure for dissolution, $p^{in} \approx E\epsilon/3$.
Inserting this pressure change into Equation (\ref{eq:total_diff}), we find 
\begin{equation}
\frac{1}{R} \frac{dR}{d \epsilon} \approx -~\frac{1}{9}~\frac{1-\phi}{\phi}~\frac{n^{sat}(T,0)}{n_L-n^{sat}(T,0)}~\frac{E}{n_Lk_BT}.
\label{eq:sensitivity_pressure}
\end{equation}
As shown by this equation, dissolution/condensation of droplets is more pronounced for stiff materials, large stretch levels, and small volume fractions of the condensed phase.

\subsection{Assessment for a specific system}

For the fluorinated-oil and silicone system explored in~\cite{Style:2018, Rosowski:2020}, $n_Lk_BT \approx 10^{6}\,\text{Pa}$,   $\frac{n^{sat}(T,0)}{n_L}\approx 3\cdot10^{-2}$, $\frac{1}{n_L}\frac{dn^{sat}(T,0)}{dT}\approx 10^{-3}\,1/(^{\circ}C)$, and $\phi_0\approx 0.04$. 
Further, we assume that $\kappa^s\approx\,0.06\,\mathrm{N/m}$, consistent with Reference~\cite{Xu:2018}.
Taking $E\approx10^3\,\mathrm{Pa}$,
\begin{equation}
    \frac{1}{R}\left|\frac{dR}{dT}\right| \approx  10^{-2}\,\frac{1}{^{\circ}C}.
\end{equation}
Thus, the droplets should change their radius by about $10\%$ for a temperature shift of $10^\circ$C.
This should be easy to resolve, and well within the range of linear response. For dissolution of droplets under applied stretch, Equation~\eqref{eq:sensitivity_pressure} gives
\begin{equation}
    \frac{1}{R}\left|\frac{dR}{d\epsilon}\right| \approx 10^{-4}.
\end{equation}
Thus, uniaxial stretches with small strains will not lead to significant dissolution.

\section{Robustness of Fitting Parameters}
\label{sec:Sensitivities}

As a final check on the viability of our approach, we investigate the ability to extract the desired material parameters from the experimental results.
We start by classifying physically admissible parameter ranges.
Surface bulk- and shear moduli are related via 
\begin{equation}
    \kappa^s=\lambda^s+\mu^s,
\end{equation}
where $\lambda^s$ is the first surface Lam\'{e} parameter. 
In two dimensions, sufficient conditions of stability for a homogenous and isotropic material require
\begin{align}
    \lambda^s+\mu^s > 0, \quad \mu^s >0, \nonumber \\
    E_{2D} >0, \quad -1 < \nu_{2D} < 1,
\end{align}
with $\nu_{2D}=\lambda^s/(\lambda^s+2\mu^s)$ \cite{Kochmann:2012}. 
Non-auxetic surfaces hence satisfy $\lambda^s\geq 0$. 
Under these conditions, five classes of parameter ranges are identified: 
\begin{align}
    \Upsilon^0& \approx (\mu^s,\kappa^s), \nonumber \\
    \Upsilon^0\ll(\mu^s,\kappa^s)&, \quad (\mu^s,\kappa^s)\ll\Upsilon^0, \nonumber \\
    \mu^s\ll(\Upsilon^0,\kappa^s)&, \quad (\Upsilon^0,\mu^s)\ll\kappa^s.
\end{align}

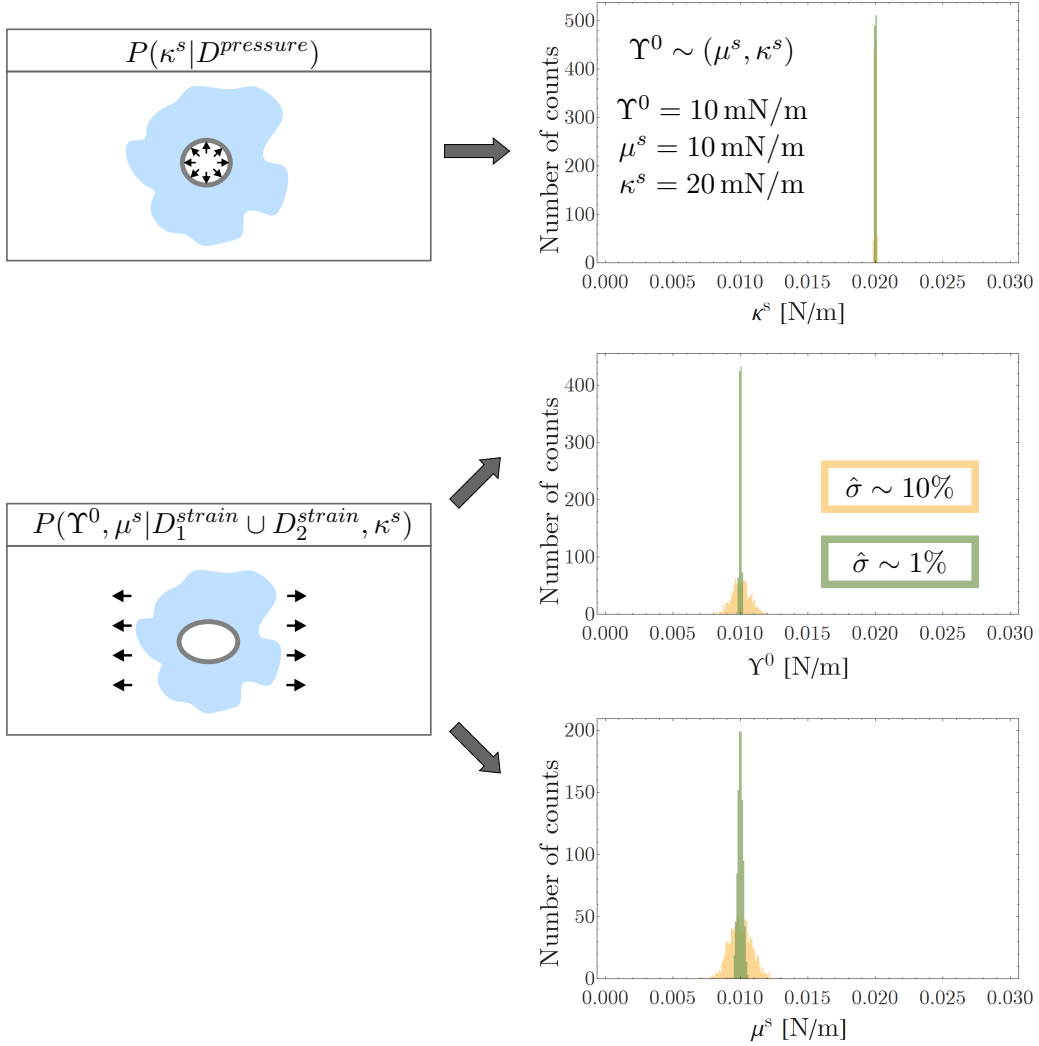
\begin{figure}[ht]
\begin{tikzpicture}
    \node (flow) at (0,0) {\includegraphics[width=0.42\textwidth]{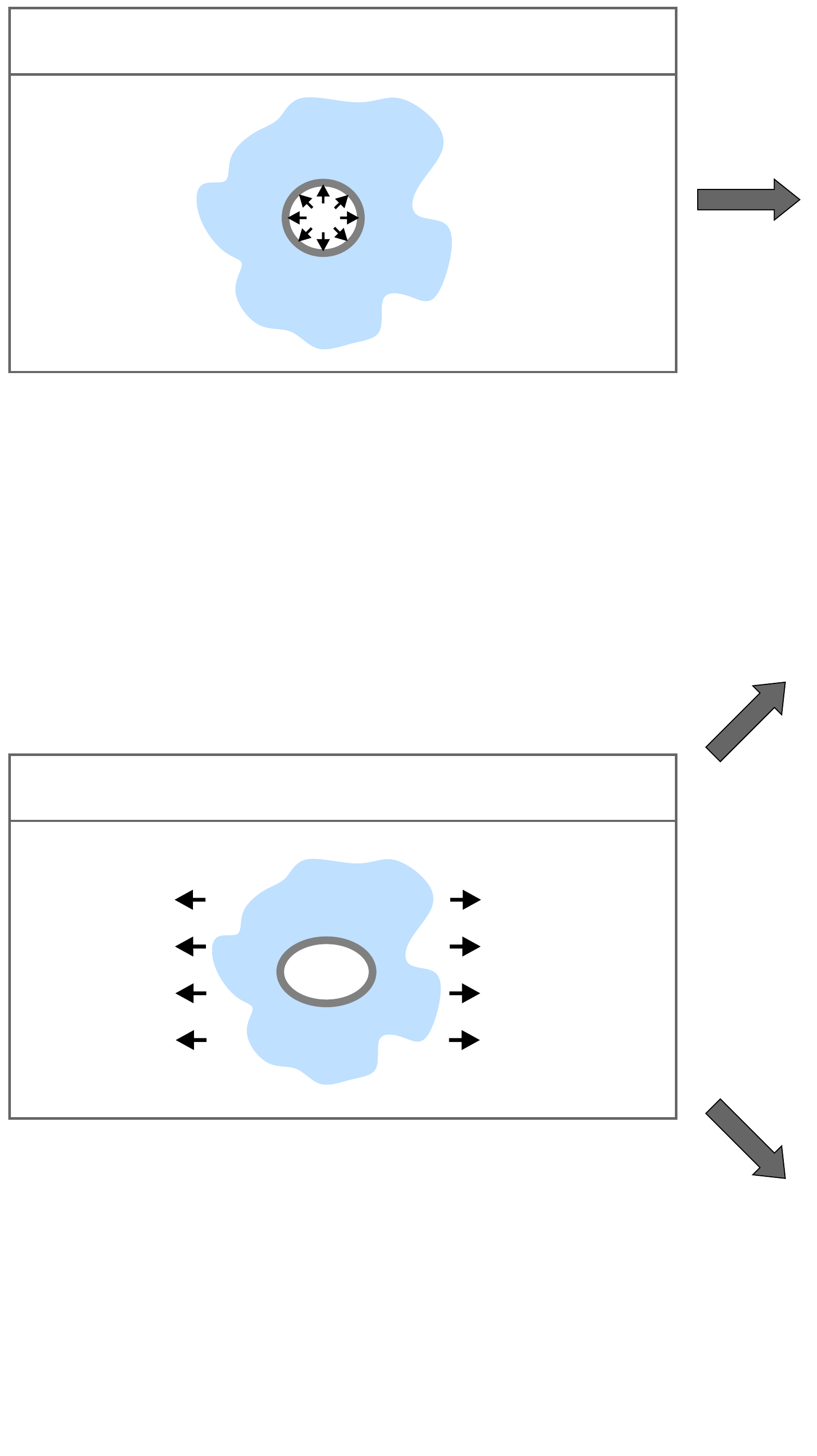}};
    \node[] at (-0.5,5.75) {$P(\kappa^s|D^{pressure})$};
    \node[] at (-0.5,-0.5) {$P(\Upsilon^0,\mu^s|D^{strain}_1 \cup D^{strain}_2,\kappa^s)$};
    \node (kappa) at (7,4.3) {\includegraphics[width=0.4\textwidth]{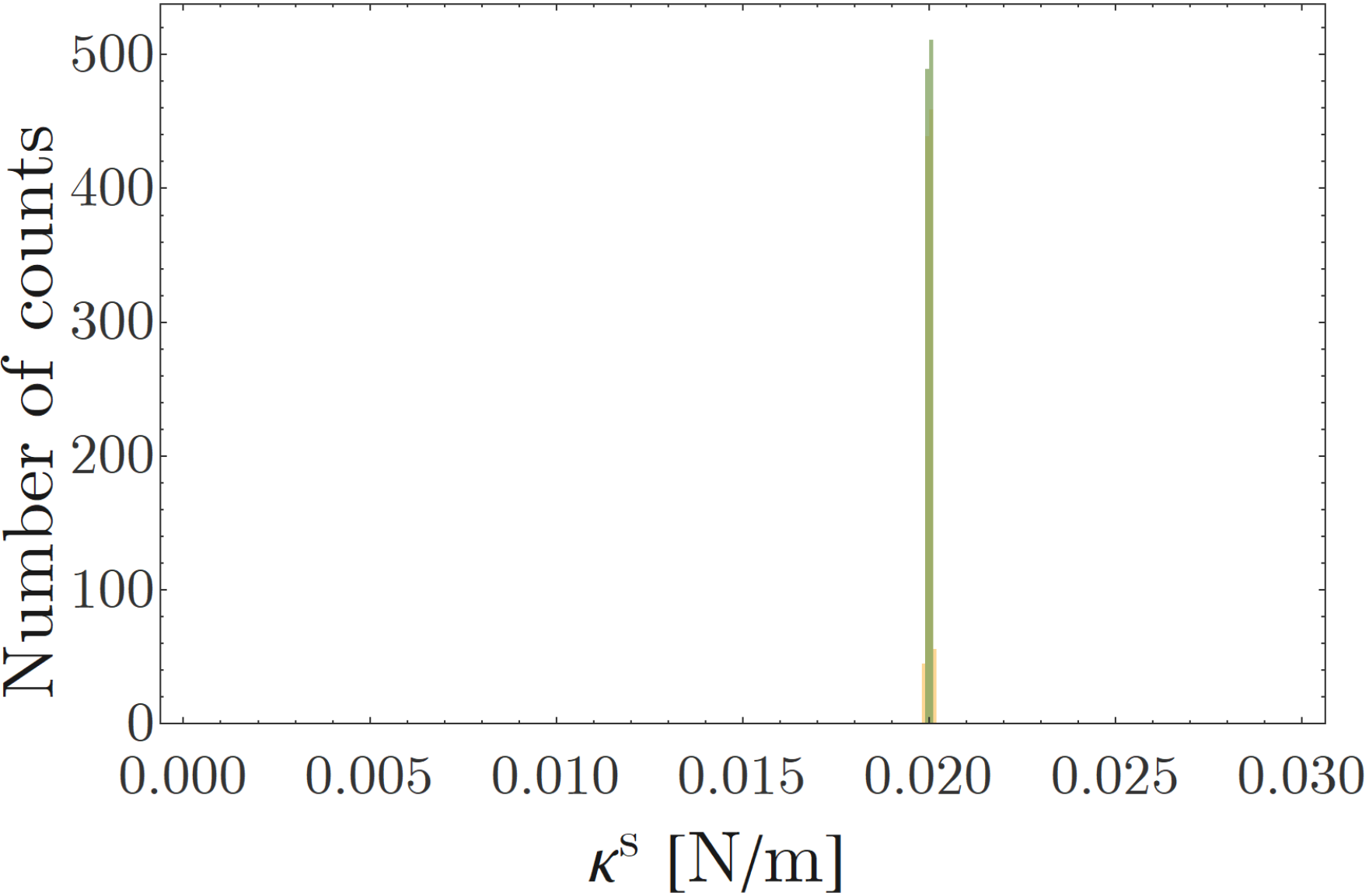}};
    \node (gamma) at (7,-0.4) {\includegraphics[width=0.4\textwidth]{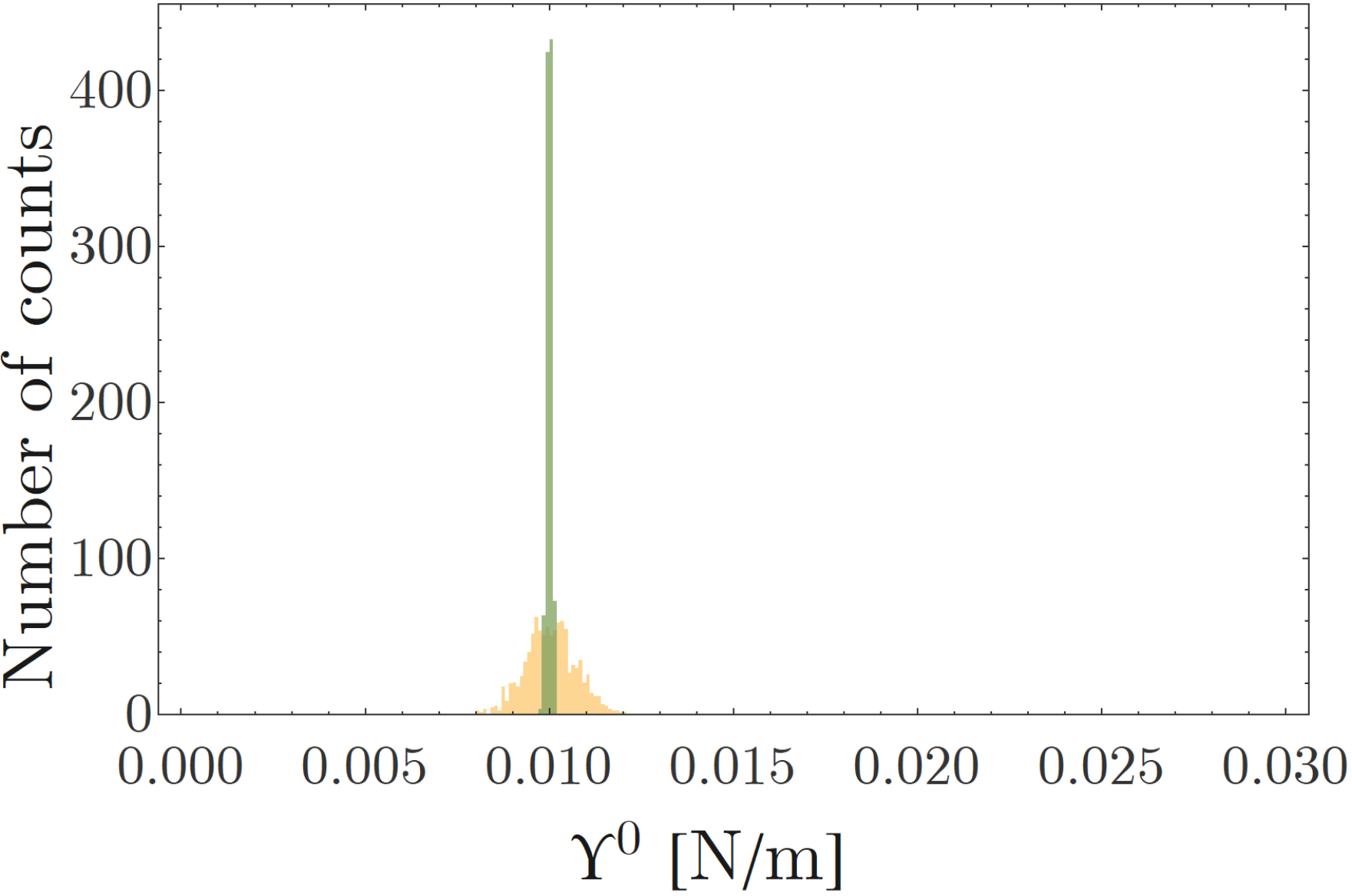}};
    \node (mu) at (7,-5.2) {\includegraphics[width=0.4\textwidth]{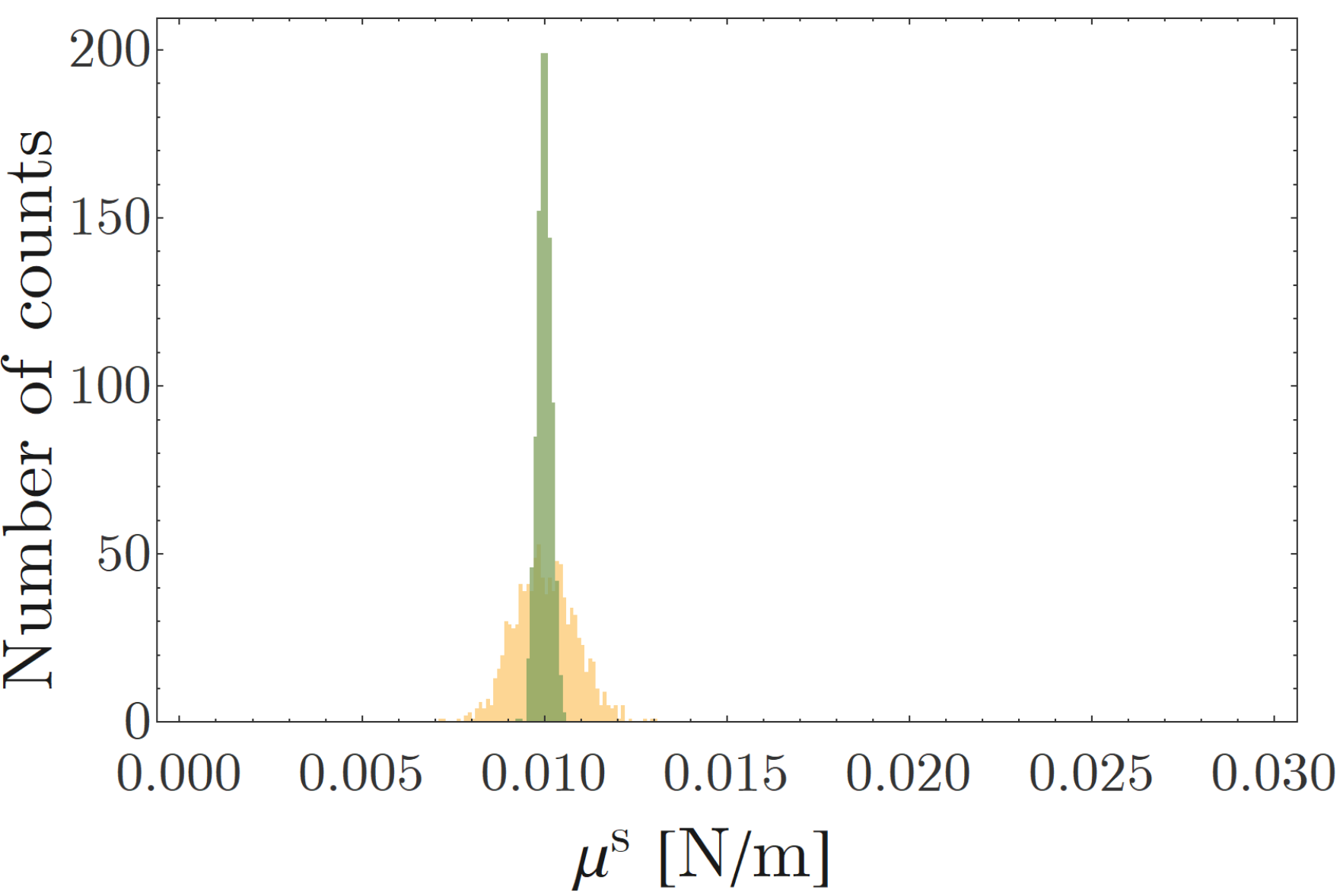}};
    \node[rectangle,
    draw = mathematica_yellow,
    line width = 1.0mm,
    minimum width = 2.0cm, 
    minimum height = 0.6cm] (r) at (8.5,0) {};
    \node[rectangle,
    draw = mathematica_green,
    line width = 1.0mm,
    minimum width = 2.0cm, 
    minimum height = 0.6cm] (r) at (8.5,-1) {};
    \node[] at (6,5.8) {$\Upsilon^0\sim(\mu^s,\kappa^s)$};
    \node[] at (6,5.) {$\Upsilon^0=10\,\text{mN/m}$};
    \node[] at (6,4.5) {$\mu^s=10\,\text{mN/m}$};
    \node[] at (6,4.) {$\kappa^s=20\,\text{mN/m}$};
    \node[] at (8.5,0.) {$\hat{\sigma}\sim 10\%$};
    \node[] at (8.5,-1) {$\hat{\sigma}\sim 1\%$};
\end{tikzpicture}
\caption{Sequential steps of the fitting procedure. An individual fit of data set $D^{pressure}$ predicts $\kappa^s$. A subsequent joint minimization over $D^{strain}_1 \cup D^{strain}_1$ determines $\Upsilon^0$ and $\mu^s$. Depicted histograms are based on surface parameters on the same order of magnitude $\Upsilon^0\sim(\mu^s,\kappa^s)$. Results for high(reduced) levels of experimental error $\hat{\sigma}$ are shown in yellow(green).}
\label{fig:Flow}
\end{figure}

We are now in place to extend the multi-modal nonlinear optimization problem first introduced in Section~\ref{sec:indi_fit}. 
With the added shape of droplets under applied hydrostatic pressure, we propose a specific minimization procedure outlined in Figure~\ref{fig:Flow}.

Our assumption for the experimental error in this analysis is based on available experimental data \cite{Style:20152}, which gives a relative standard deviation $\sim 12\%$. 
With further optimization, the experimental error could likely be reduced to $\sim 1\%$.
Here, we examine both levels of experimental noise and the resulting optimization sensitivities.

The proposed procedure requires an initial minimization over the individual distribution $P(\kappa^s| D^{pressure})$, where we exploit that the shape of droplets under hydrostatic pressure solely depends on the surface bulk modulus $\kappa^s$.
Initial droplet radii are recorded and final droplet radii after growth/shrinkage as well as the temperature difference $\Delta T$ are measured.
Experimental data is fitted using Equation~\eqref{eq:governing_sym}, with inflation pressure calculated from~\eqref{eq:p}.
Figure~\ref{fig:Flow} shows that $\kappa^s$ can be recovered for both levels of experimental noise.

We then turn to the joint probability distribution of $(\Upsilon^0,\mu^s)$ conditional on $\kappa^s$, \\ $P(\Upsilon^0,\mu^s|D^{strain}_1 \cup D^{strain}_2,\kappa^s)$. 
Initial droplet radii are recorded and the attained final length- and width of droplets at fixed uniaxial far-field strain is measured.
Using axisymmetric loading conditions, hoop displacements are determined at distinct points initially located at $\theta=\theta_0$ (with $\theta_0$ given in Equation~\eqref{eq:theta}) . 
As sketched in Figure~\ref{fig:Construction}, the distance within the hoop direction $\mathbf{g}_{\theta}$ to deformed points at the ellipsoidal surface is measured.  Experimental data is fitted using Equations~\eqref{eq:drop_strain} and~\eqref{eq:hoop_disp}.

As illustrated in Figure~\ref{fig:Flow}, a spectrum of narrow width can be recovered in both $\Upsilon^0$ and $\mu^s$ even at high levels of experimental noise, whereby no systematic shifts are encountered.
Test parameters at reduced experimental error $\hat{\sigma}\sim 1\%$ are recovered at high precision. 

In the examined test case, all surface parameters are on the same order of magnitude, $\Upsilon^0\sim(\mu^s,\kappa^s)$.
This case serves as the worst-case scenario on the resultant fitting accuracy.
For all remaining physically admissible parameter ranges, fitting accuracies are enhanced, as shown in Section~\ref{sec:sensitivities_appendix}.
For the different parameter ranges illustrated in Figures~\ref{fig:Histo_Appendix} and~\ref{fig:Histo_Appendix_2}, we focus on the estimate of ($\Upsilon^0,\mu^s$), which constitutes the only multi-modal and hence most sensitive minimization problem.
Finally, it bears mentioning that a reduced experimental error does not qualitatively change the necessity for the outlined minimization procedure. 

\section{Conclusions}

The quantification of surface elasticity in soft solids calls for a careful distinction between surface- and bulk effects. 
Here, we derive analytical solutions for two modes of deformation which are readily accessible in a laboratory setting and allow for two key advantages:
Both deformations are amenable to be employed in the small deformation regime, hence circumventing possible material nonlinearities.
Droplets furthermore feature geometrically smooth surfaces to avoid stress divergences within the bulk material.
In comparison to recent work on the calibration of surface hyperelastic constitutive models \emph{via} mechanical tests on a soft cylindrical specimen \cite{Basu:2021}, the proposed experiments allow for an easy fabrication as well as stable geometries.
In addition, experiments systematically yield a population of droplets, such that a distribution of droplet sizes is obtained.
The derived modes of deformation are coupled to the dissolution of droplets.
It is shown that droplet dissolution/ condensation is more pronounced for stiff materials, large stretch levels, and small volume fractions of the condensed phase.
We find that a robust quantification of the complete set of isotropic surface material parameters $(\Upsilon^0,\mu^s,\kappa^s)$ requires to take into account both modes of deformation. 
By following the outlined minimization procedure, surface elastic constants can be reliably recovered at realistic levels of experimental noise, even for 
a worst-case scenario in which all surface parameters are on the same order of magnitude, $\Upsilon^0\sim(\mu^s,\kappa^s)$.

We hope that this new approach to measuring surface elastic constants of soft solids will resolve the controversy surrounding this topic, and enable systematic studies to understand the structure-property relationships that underly surface elasticity.

\section*{Acknowledgements}
This work was funded via the \emph{SNF Ambizione} grant PZ00P$2186041$. Great thanks goes to Robert Style (ETH Zurich) and Nicolas Bain (ETH Zurich) for many helpful discussions. We thank Siddhant Kumar (TU Delft) for his advice on stochastics of multivariate optimizations problems, as well as Pradeep Sharma (University of Houston) for sharing his expertise in the realm of surface elasticity. On `surface' matters in a more abstract sense, we enjoyed discussions on greek philology with Nicola Schmid (University of Zurich).

\appendix

\section{Differential surface Operators}
\label{sec:Surface_Operators}

The surface projection tensor utilized to project second order tensors onto curved surfaces is defined as
\begin{equation}
    \mathbb{P}(\mathbf{X}) = \mathbf{I} - \mathbf{N}(\mathbf{X})\otimes\mathbf{N}(\mathbf{X}),
\end{equation}
where $\mathbf{I}$ is the identity tensor and $\mathbf{N}(\mathbf{X})$ is the outward unit normal to the tangent surface at $\mathbf{X}$. Let $\mathbf{T}(\mathbf{X})$ be a smooth second order tensor field and $\mathbf{t}(\mathbf{X})$ a smooth vector field. On a smooth surface with ourward unit normal $\mathbf{N}(\mathbf{X})$, their projections are given as
\begin{equation}
    \mathbf{T}^s = \mathbb{P}\mathbf{T}\mathbb{P}, \quad \text{and} \quad \mathbf{t}^s = \mathbb{P}\mathbf{t}. 
\end{equation}
The surface identity tensor is hence obtained as
\begin{equation}
    \mathbf{I}^s = \mathbb{P}\mathbf{I}\mathbb{P}.
\end{equation}
Surface gradients of a smooth scalar field $t(\mathbf{X}):\Omega_0\rightarrow\mathbb{R}$ and a smooth vector field $\mathbf{t}(\mathbf{X}):\Omega_0\rightarrow\mathbb{R}^3$ follow from
\begin{equation}
    \boldsymbol{\nabla}_st = \mathbb{P}\boldsymbol{\nabla}t, \quad \text{and} \quad \boldsymbol{\nabla}_s\mathbf{t} = (\boldsymbol{\nabla}\mathbf{t})\mathbb{P}.
\end{equation}
In addition, the surface divergence of a smooth vector field $\mathbf{t}(\mathbf{X})$ and a smooth second order tensor field $\mathbf{T}(\mathbf{X})$ are 
\begin{equation}
    \text{div}_s(\mathbf{t}) = \text{tr}(\boldsymbol{\nabla}_s\mathbf{t}) \quad \text{and} \quad \mathbf{v}\cdot\text{div}_s(\mathbf{T}) = \text{div}_s(\mathbf{T}^T\mathbf{v}),
\end{equation}
with $\mathbf{v}\in\mathbb{R}^3$ being an arbitrary constant vector. In a more general case, we have
\begin{equation}
    \mathbf{N}\cdot\text{div}_s(\mathbf{T}) = \text{div}_s(\mathbf{T}^T\mathbf{N}) - \mathbf{T}\cdot\boldsymbol{\nabla}_s\mathbf{N}.
\label{eq:surf_div}
\end{equation}

\section{Equivalence between linear elasticity and Stokes flow of Newtonian fluids}
\label{sec:Equiv}

The static equilibrium of an elastic body and steady Stokes flows are described by equivalent governing equations.
Assuming linearized kinematics, strain tensor $\boldsymbol{\epsilon}$ and displacements $\mathbf{u}$ are related as 
\begin{equation}
    \boldsymbol{\epsilon} = \frac{1}{2}\left(\boldsymbol{\nabla}\mathbf{u}+\boldsymbol{\nabla}\mathbf{u}^T\right).
\end{equation}
For linear elastic material behavior, stresses follow as
\begin{equation}
    \boldsymbol{\sigma} = 2\mu\boldsymbol{\epsilon} + \left(\kappa-\frac{2}{3}\mu\right)\text{tr}(\boldsymbol{\epsilon})\mathbf{I},
\end{equation}
where $\mu$ is the shear modulus and $\kappa$ is the bulk modulus.
Insertion into the momentum balance in combination with the condition of incompressibility $\text{tr}(\boldsymbol{\epsilon})=0$ gives
\begin{equation}
    -\boldsymbol{\nabla}p + \mu\boldsymbol{\nabla}^2\mathbf{u} = \mathbf{0},
\label{eq:final_solid}
\end{equation}
where $p$ is the hydrostatic pressure term $p=-\text{tr}(\boldsymbol{\sigma})$.

For Newtonian fluids, similar to linear elasticity, viscous stresses arising from the flow field are linearly related to the local rate-of-strain-tensor
\begin{equation}
    \mathbf{D} = \frac{1}{2}\left(\boldsymbol{\nabla}\mathbf{v}+\boldsymbol{\nabla}\mathbf{v}^T\right),
\end{equation}
such that
\begin{equation}
    \boldsymbol{\sigma} = 2\eta\mathbf{D} + \left(\zeta-\frac{2}{3}\eta\right)\text{tr}(\mathbf{D})\mathbf{I},
\end{equation}
where $\zeta$ is the bulk viscosity and $\eta$ is the shear viscosity.
Insertion into the momentum balance in combination with an incompressible, divergence-free flow field $\text{div}(\mathbf{v})=0$ again gives Equation~\eqref{eq:final_solid}, using the substitutions $\eta=\mu$ and $\mathbf{v}=\mathbf{u}$.
This illustrates the equivalence of linear elastic statics and steady Stokes flow of Newtonian fluids.
Solutions to Stoke's flow in spherical geometries incorporating elastic interfaces are well developed \cite{Biesel:1981,Biesel:1985}. Here, we adopt them for the solution of the volume preserving elastic problem as outlined below.

\section{Fundamental set of solutions of the Stokes equations}
\label{sec:fundamental_solutions}

The velocity basis functions are 
\begin{subequations}
\label{vel basis -}
\begin{align}
\label{-vel 0}
		\bv^-_{jm0}&={\textstyle\frac{1}{2}}r^{-j}\left(2-j+j r^{-2}\right)\by_{jm0}
+{\textstyle\frac{1}{2}}r^{-j}\left[j\left(j+1\right)\right]^{\frac{1}{2}}\left(
1-r^{-2}\right) \by_{jm2} \, , \\
\label{-vel 1}
		\bv^-_{jm1}&=\textstyle r^{(-j-1)} \by_{jm1} \, , \\
\label{-vel 2}
		\bv^-_{jm2}&={\textstyle\frac{1}{2}}r^{-j}\left(2-j\right)
(\textstyle\frac{j}{j+1})^{\frac{1}{2}}\left(1-r^{-2}\right)\by_{jm0}
		+{\textstyle\frac{1}{2}}r^{-j}\left(j+(2-j)r^{-2}\right)\by_{jm2} \, ,
\end{align}
\end{subequations}
\vspace{-3ex}
\begin{subequations}
\label{vel basis +}
\begin{align}
\label{+vel 0}
		\bv^+_{jm0}&= {\textstyle\frac{1}{2}}r^{j-1}\left(-(j+1)+(j
+3)r^2\right)\by_{jm0} 
-{\textstyle\frac{1}{2}}r^{j-1}\left[j\left(j+1\right)\right]^{\frac{1}{2}}\left
(1-r^2\right)\by_{jm2} \, , \\
\label{+vel 1}
		\bv^+_{jm1}&=\textstyle r^j \by_{jm1} \, , \\
\label{+vel 2}
\bv^+_{jm2}&={\textstyle\frac{1}{2}}r^{j-1}\left(3+j\right)(\textstyle\frac{j+1}
{j})^{\frac{1}{2}}\left(1-r^2\right)\by_{jm0}
		+{\textstyle\frac{1}{2}}r^{j-1}\left(j +3-(j+1)r^2  \right)\by_{jm2} \, .
\end{align}
\end{subequations}
On a sphere $r=1$ these velocity fields reduce to the vector spherical harmonics defined by \refeq{vector_harmonics_1}
\begin{equation}
\bv^{\pm}_{jmq}=\bS_{jmq}\,.
\end{equation}
The  tractions on a sphere due to the displacement fields \refeq{vel basis -} and \refeq{vel basis +} are 
\begin{subequations}
\label{HD trac:1}
\begin{equation}
\tau^{+}_{jm0}=(2j+1)c^+_{jm0}-3\left(\frac{j+1}{j}\right)^\half c^+_{jm2}\,, 
\end{equation}
\begin{equation}
 \tau^{-}_{jm0}=-(2j+1)c^-_{jm0}+3\left(\frac{j}{j+1}\right)^\half c^-_{jm2}\,
 \end{equation}
\end{subequations}

\begin{subequations}
\label{HD trac:2}
\begin{equation}
\tau^{-}_{jm2}=3\left(\frac{j}{j+1}\right)^\half c^-_{jm0}-\frac{4+3j+2j^2}{j+1} c^-_{jm2}\,, 
\end{equation}
\begin{equation}
\tau^{+}_{jm2}=-3\left(\frac{j+1}{j}\right)^\half c^+_{jm0}+\frac{3+j+2j^2}{j} c^+_{jm2}\,
 \end{equation}
\end{subequations}

\begin{equation}
\label{HD trac:C}
\begin{split}
\tau^{-}_{jm1} =-(j+2)c^-_{jm1}\,, \quad 
\tau^{+}_{jm1} =(j-1)c^+_{jm1}\,,
\end{split}
\end{equation}

\section{Spherical harmonics}
\label{Harmonics}

An extensive reference on spherical harmonics is a handbook by Varshalovich et al. \cite{Varshalovich:1988}. Their properties in relation to the problems of particle microhydrodynamics are summarized in  Blawzdziewicz et al. \cite{Blawzdziewicz-Vlahovska-Loewenberg:2000}.

The normalized spherical scalar harmonics are defined as
\begin{equation}
\label{normalized spherical harmonics}
   Y_{jm}\left(\theta,\varphi\right) = \textstyle \left[\frac{2j+1}{4\pi}\frac{(j-m)!}{(j+m)!}\right]^\half (-1)^m P_j^m(\cos\theta)e^{{\rm i}m\varphi},
\end{equation}
where  $\rhat=\br/r$,  $(r, \theta,\varphi)$
are the spherical coordinates, and $P_j^m(\cos\theta)$ are the Legendre polynomials.
The vector spherical harmonics 
are defined as
\begin{equation}
\label{vector_harmonics_1}
\begin{split}
\bS_{jm0}&=\left[j\left(j+1\right)\right]^{-\half}r\nabla_\Omega Y_{jm}\,,\quad \bS_{jm2}=\rhat Y_{jm}\,,\quad \bS_{jm1}=-\im \rhat \times \bS_{jm0}
\end{split}
\end{equation}
where $\nabla_{\Omega}$ denotes the angular part of the gradient operator. In spherical coordinates,  the vector spherical harmonics that are tangential to a sphere are
\begin{equation}
\label{vector_harmonics_2}
\begin{split}
\bS_{jm0}&=\frac{1}{\sqrt{j(j+1)}} \frac{\partial Y_{jm}}{\partial \theta}  \that+\frac{\im m}{\sqrt{j(j+1)}} \frac{Y_{jm}}{\sin \theta}\phihat\\
\bS_{jm1}&=-\frac{ m}{\sqrt{j(j+1)}} \frac{Y_{jm}}{\sin \theta}\that-\frac{\im }{\sqrt{j(j+1)}} \frac{\partial Y_{jm}}{\partial \theta}  \phihat\\
\end{split}
\end{equation}
For example
\begin{equation}
\label{vh2}
\bS_{200}=-\textstyle\sqrt{\frac{15}{32\pi}} \sin(2 \theta)\ehat_\theta\,,\quad \bS_{101}= \textstyle -\im \sqrt{\frac{3}{8\pi}} \sin\theta \ehat_\varphi\,,\quad
\bS_{202}=\textstyle \frac{1}{8}\sqrt{\frac{5}{\pi}}[1+3\cos(2\theta)]\rhat\,.
\end{equation}

\section{Interfacial displacement field}
\label{sec:disp_appendix}

The interfacial displacement field is given as
\begin{align}
    \mathbf{u} &= \frac{5e^{-2i\varphi}ER_0^2(2ER_0+9\kappa^s+3\mu^s)(e^{2i\varphi}c_z+3e^{2i\varphi}c_z\cos{2\theta}+(1+e^{4i\varphi})(c_x-c_y)\sin{\theta}^2)}{12\left(2E^2R_0^2+12\mu^s\kappa^s+6\Upsilon^0(3\kappa^s+2\mu^s)+ER_0(5\Upsilon^0+8\kappa^s+6\mu^s)\right)}\mathbf{g}_r \nonumber \\
    &- \frac{5ER_0^2(2ER_0+3\Upsilon^0+6\kappa^s)(3c_z+(c_y-c_x)\cos{2\varphi})\sin{2\theta}}{12\left(2E^2R_0^2+12\mu^s\kappa^s+6\Upsilon^0(3\kappa^s+2\mu^s)+ER_0(5\Upsilon^0+8\kappa^s+6\mu^s)\right)}\mathbf{g}_{\theta} \nonumber \\
    &- \frac{5ER_0^2(c_x-c_y)(2ER_0+3\Upsilon^0+6\kappa^s)\sin{\theta}\sin{2\varphi}}{6\left(2E^2R_0^2+12\mu^s\kappa^s+6\Upsilon^0(3\kappa^s+2\mu^s)+ER_0(5\Upsilon^0+8\kappa^s+6\mu^s)\right)}\mathbf{g}_{\varphi}.
\label{eq:disp_r_appendix}
\end{align}

\section{Fitting sensitivities}
\label{sec:sensitivities_appendix}

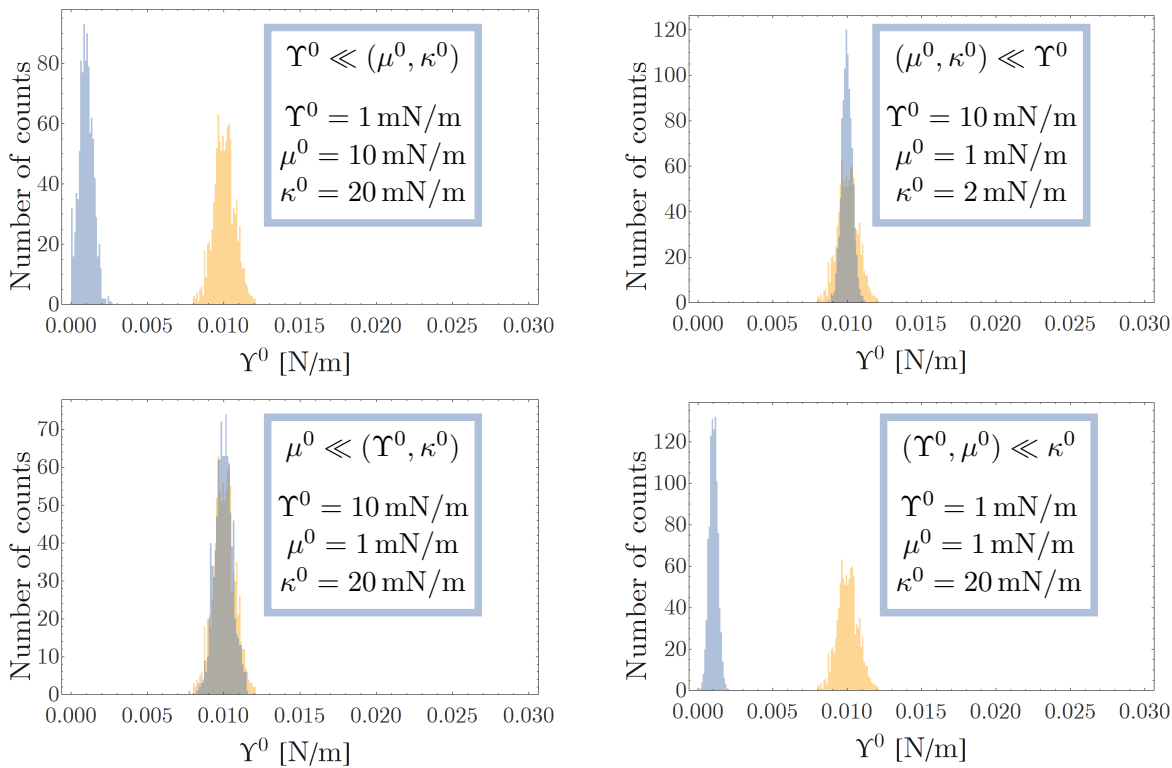
\begin{figure}[ht]
\begin{subfigure}{.49\textwidth}
  \centering
  \begin{tikzpicture}
      \node (histo_1) at (0,0) {\includegraphics[width=0.9\textwidth]{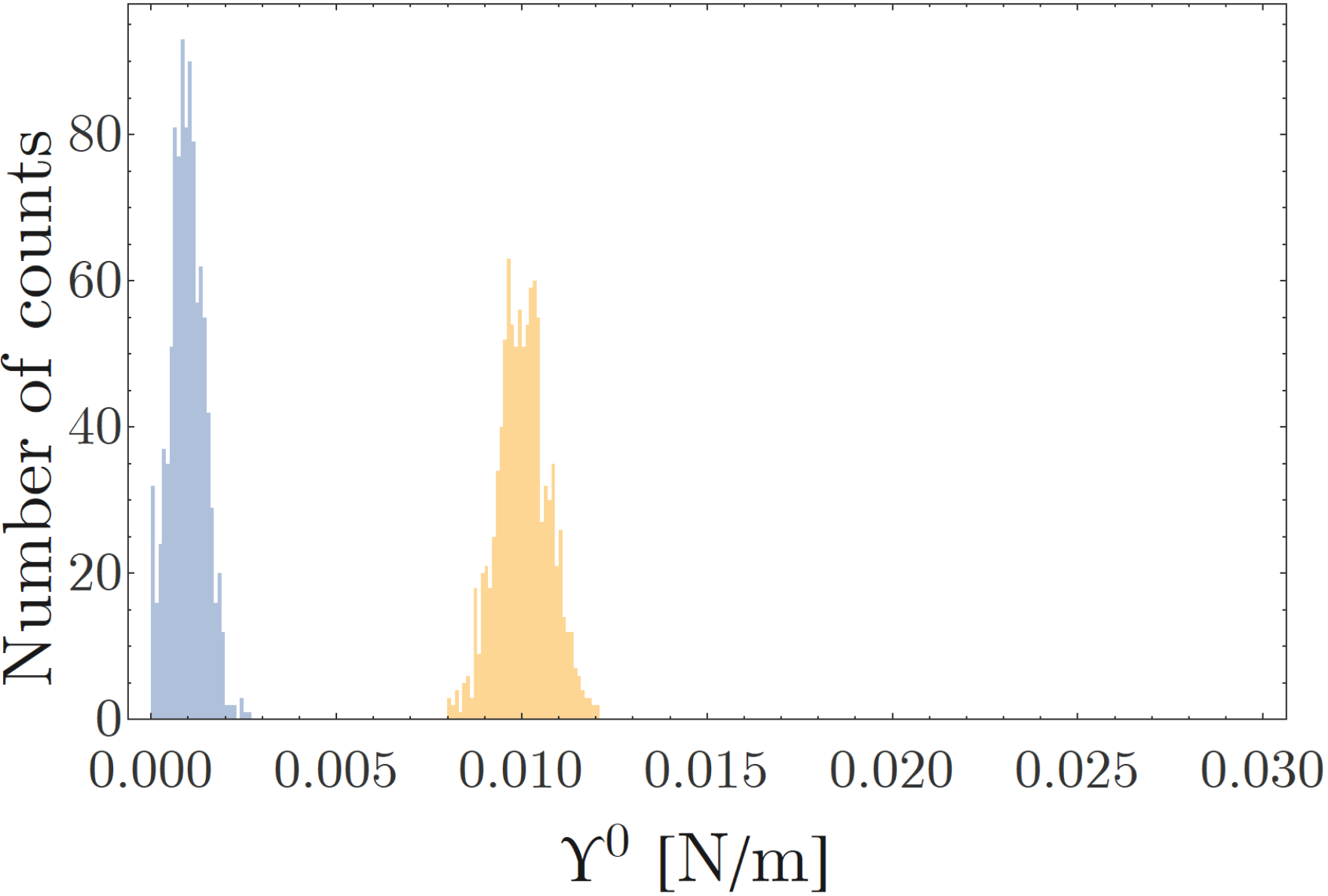}};
      \node[rectangle,
    draw = mathematica_blue,
    line width = 1.0mm,
    minimum width = 2.8cm, 
    minimum height = 2.6cm] (r) at (1.2,0.9) {};
    \node[] at (1.2,1.8) {$\Upsilon^0\ll(\mu^0,\kappa^0)$};
    \node[] at (1.2,1) {$\Upsilon^0=1\,\text{mN/m}$};
    \node[] at (1.2,0.5) {$\mu^0=10\,\text{mN/m}$};
    \node[] at (1.2,0) {$\kappa^0=20\,\text{mN/m}$};
  \end{tikzpicture}
\end{subfigure}
\begin{subfigure}{.49\textwidth}
  \centering
  \begin{tikzpicture}
      \node (histo_2) at (0,0) {\includegraphics[width=0.9\textwidth]{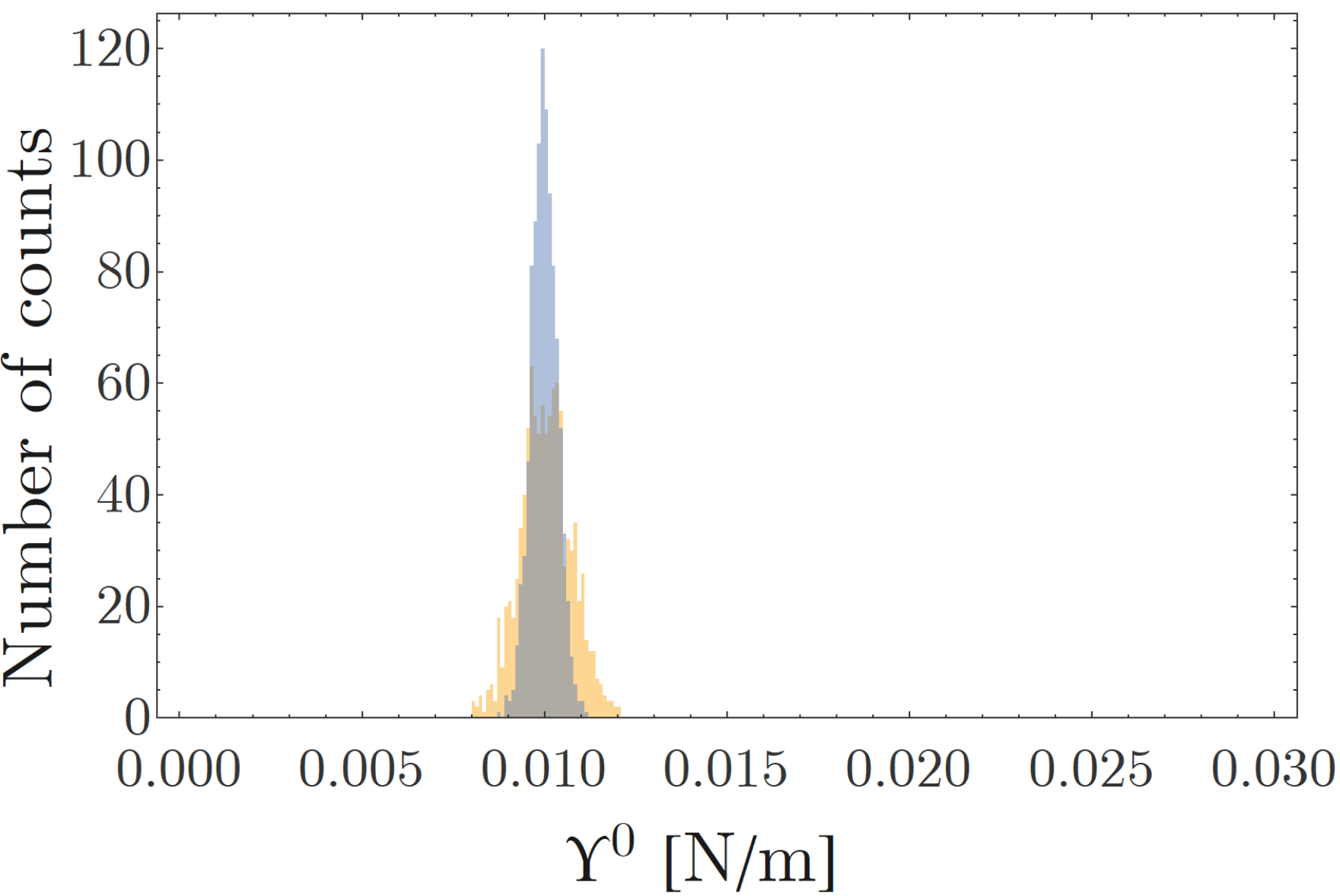}};
    \node[rectangle,
    draw = mathematica_blue,
    line width = 1.0mm,
    minimum width = 2.8cm, 
    minimum height = 2.6cm] (r) at (1.1,0.9) {};
    \node[] at (1.1,1.8) {$(\mu^0,\kappa^0)\ll\Upsilon^0$};
    \node[] at (1.1,1) {$\Upsilon^0=10\,\text{mN/m}$};
    \node[] at (1.1,0.5) {$\mu^0=1\,\text{mN/m}$};
    \node[] at (1.1,0) {$\kappa^0=2\,\text{mN/m}$};
  \end{tikzpicture} 
\end{subfigure}
\begin{subfigure}{.49\textwidth}
  \centering
  \begin{tikzpicture}
      \node (histo_3) at (0,0) {\includegraphics[width=0.9\textwidth]{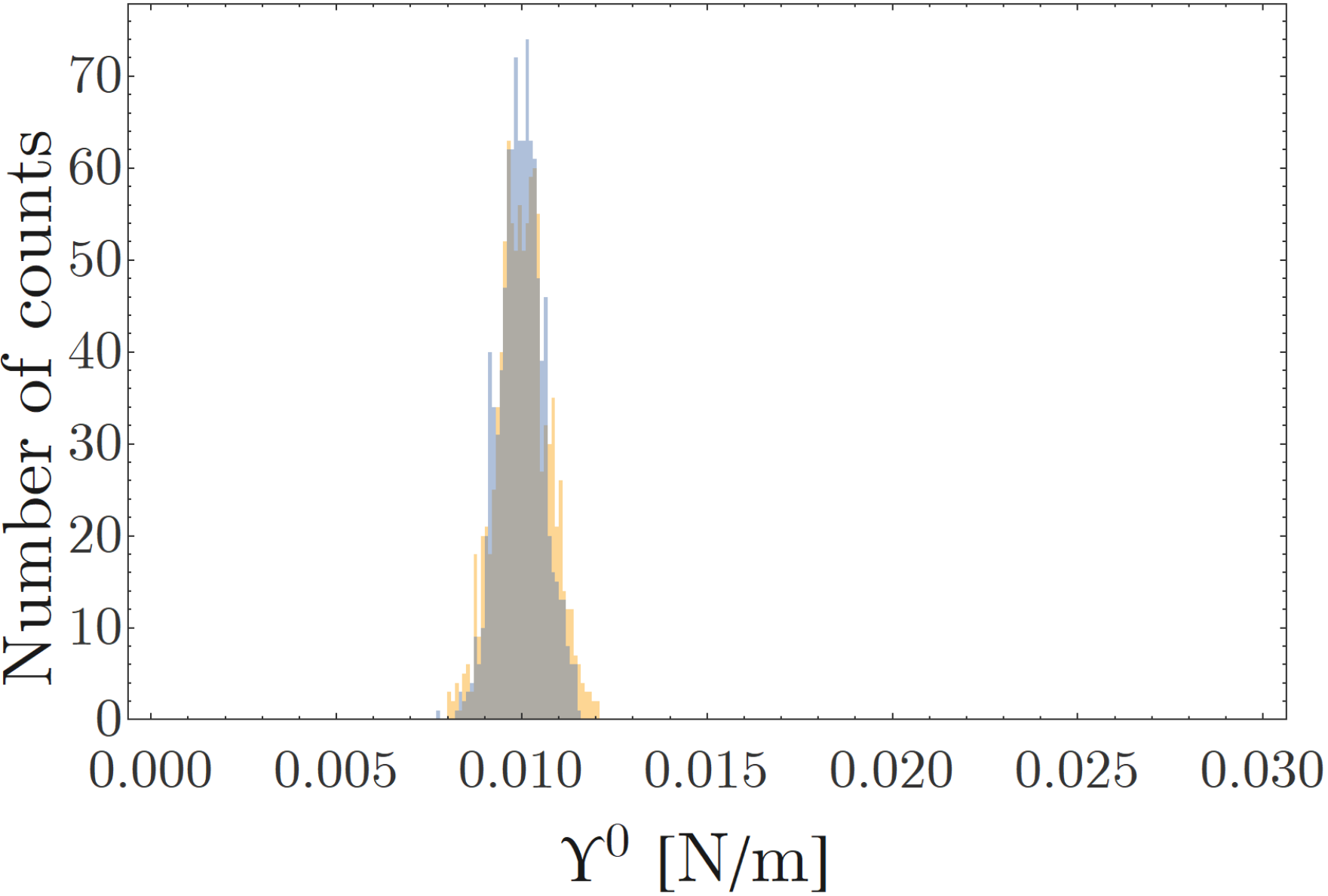}};
      \node[rectangle,
    draw = mathematica_blue,
    line width = 1.0mm,
    minimum width = 2.8cm, 
    minimum height = 2.6cm] (r) at (1.2,0.9) {};
      \node[] at (1.2,1.8) {$\mu^0\ll(\Upsilon^0,\kappa^0)$};
      \node[] at (1.2,1.) {$\Upsilon^0=10\,\text{mN/m}$};
      \node[] at (1.2,0.5) {$\mu^0=1\,\text{mN/m}$};
      \node[] at (1.2,0) {$\kappa^0=20\,\text{mN/m}$};
  \end{tikzpicture}  
\end{subfigure}
\begin{subfigure}{.49\textwidth}
  \centering
  \begin{tikzpicture}
      \node (histo_4) at (0,0) {\includegraphics[width=0.9\textwidth]{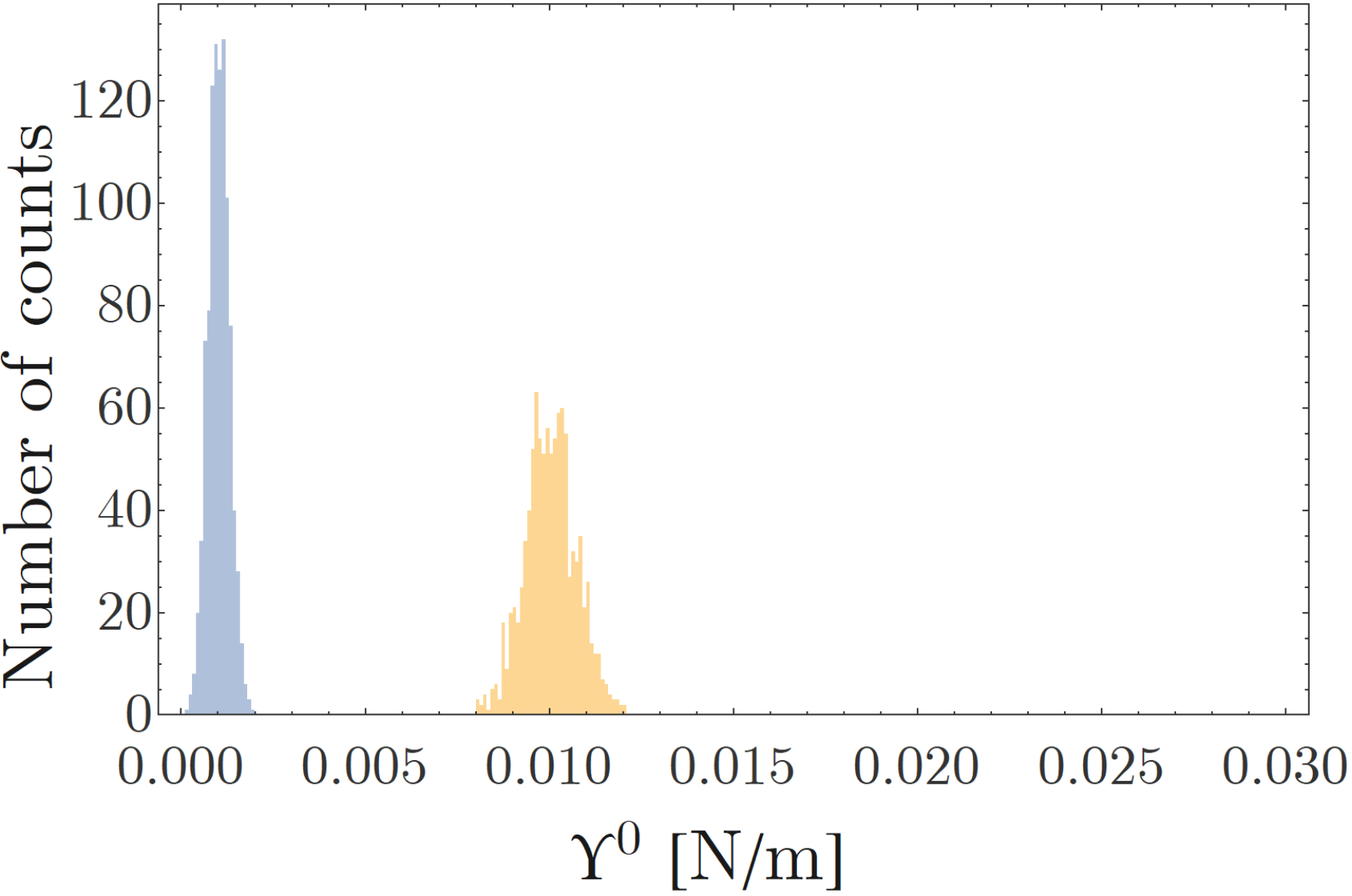}};
      \node[rectangle,
    draw = mathematica_blue,
    line width = 1.0mm,
    minimum width = 2.8cm, 
    minimum height = 2.6cm] (r) at (1.2,0.9) {};
    \node[] at (1.2,1.8) {$(\Upsilon^0,\mu^0)\ll\kappa^0$};
    \node[] at (1.2,1) {$\Upsilon^0=1\,\text{mN/m}$};
    \node[] at (1.2,0.5) {$\mu^0=1\,\text{mN/m}$};
    \node[] at (1.2,0) {$\kappa^0=20\,\text{mN/m}$};
  \end{tikzpicture} 
\end{subfigure}
\caption{Fitting sensitivities of $\Upsilon^0$ for varying parameter ranges (blue) in comparison to results for $\Upsilon^0\sim(\mu^s,\kappa^s)$ (yellow), both at high levels of experimental error $\hat{\sigma}\sim 10\%$.}
\label{fig:Histo_Appendix}
\end{figure}

\begin{figure}[ht]
\begin{subfigure}{.49\textwidth}
  \centering
  \begin{tikzpicture}
      \node (histo_1) at (0,0) {\includegraphics[width=0.9\textwidth]{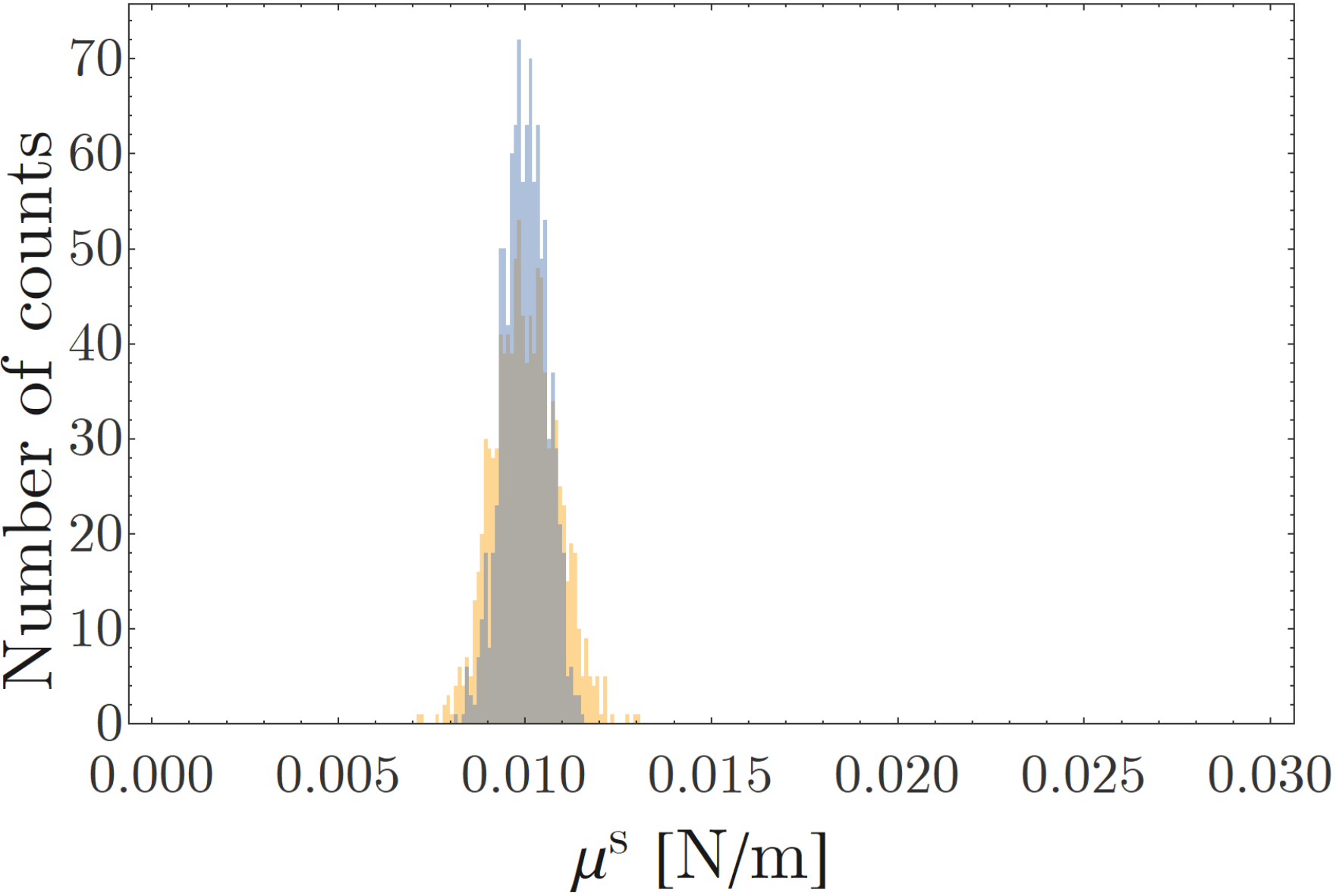}};
      \node[rectangle,
    draw = mathematica_blue,
    line width = 1.0mm,
    minimum width = 2.8cm, 
    minimum height = 2.6cm] (r) at (1.2,0.9) {};
    \node[] at (1.2,1.8) {$\Upsilon^0\ll(\mu^0,\kappa^0)$};
    \node[] at (1.2,1) {$\Upsilon^0=1\,\text{mN/m}$};
    \node[] at (1.2,0.5) {$\mu^0=10\,\text{mN/m}$};
    \node[] at (1.2,0) {$\kappa^0=20\,\text{mN/m}$};
  \end{tikzpicture}
\end{subfigure}
\begin{subfigure}{.49\textwidth}
  \centering
  \begin{tikzpicture}
      \node (histo_2) at (0,0) {\includegraphics[width=0.9\textwidth]{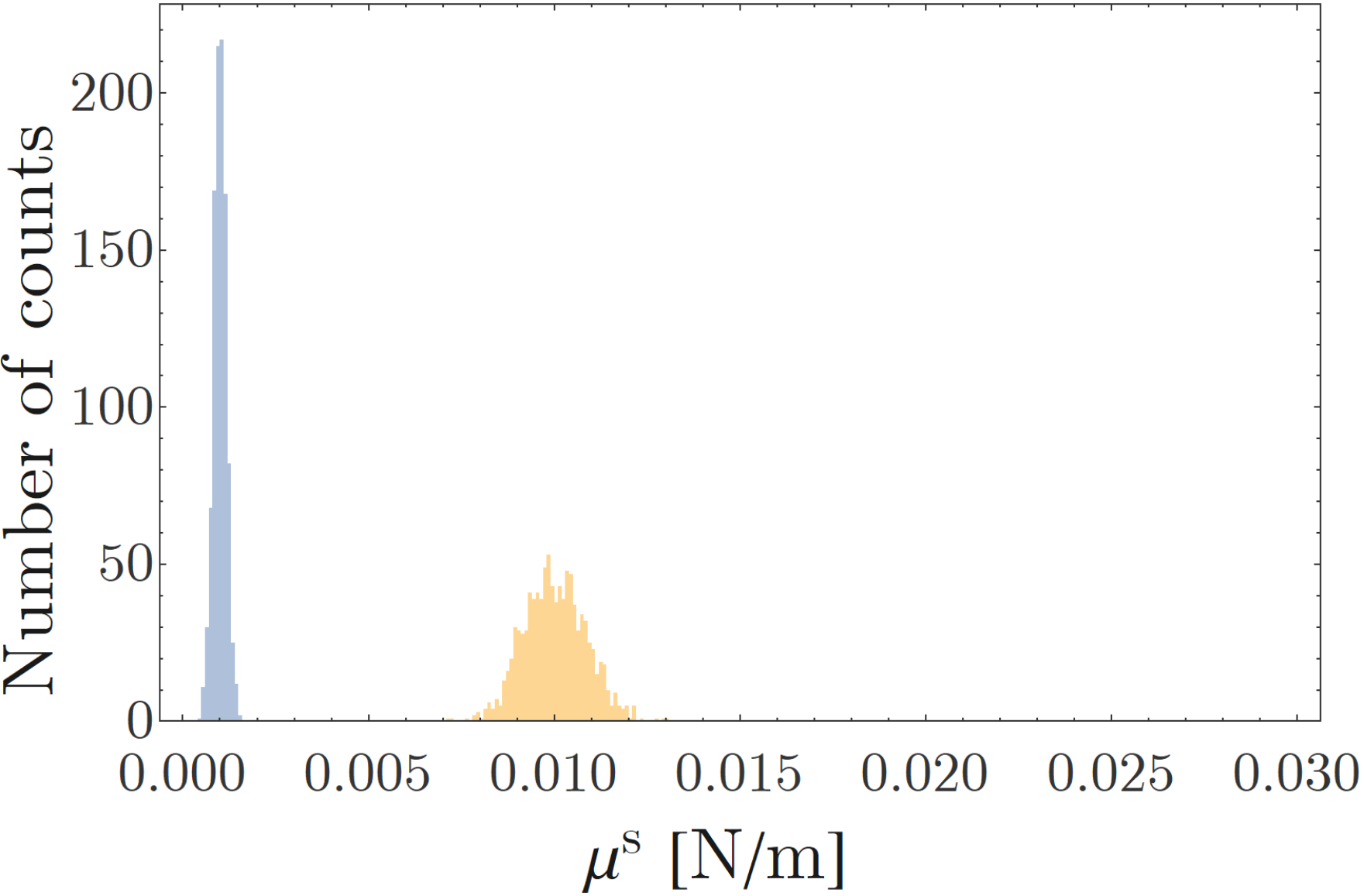}};
    \node[rectangle,
    draw = mathematica_blue,
    line width = 1.0mm,
    minimum width = 2.8cm, 
    minimum height = 2.6cm] (r) at (1.1,0.9) {};
    \node[] at (1.1,1.8) {$(\mu^0,\kappa^0)\ll\Upsilon^0$};
    \node[] at (1.1,1) {$\Upsilon^0=10\,\text{mN/m}$};
    \node[] at (1.1,0.5) {$\mu^0=1\,\text{mN/m}$};
    \node[] at (1.1,0) {$\kappa^0=2\,\text{mN/m}$};
  \end{tikzpicture} 
\end{subfigure}
\begin{subfigure}{.49\textwidth}
  \centering
  \begin{tikzpicture}
      \node (histo_3) at (0,0) {\includegraphics[width=0.9\textwidth]{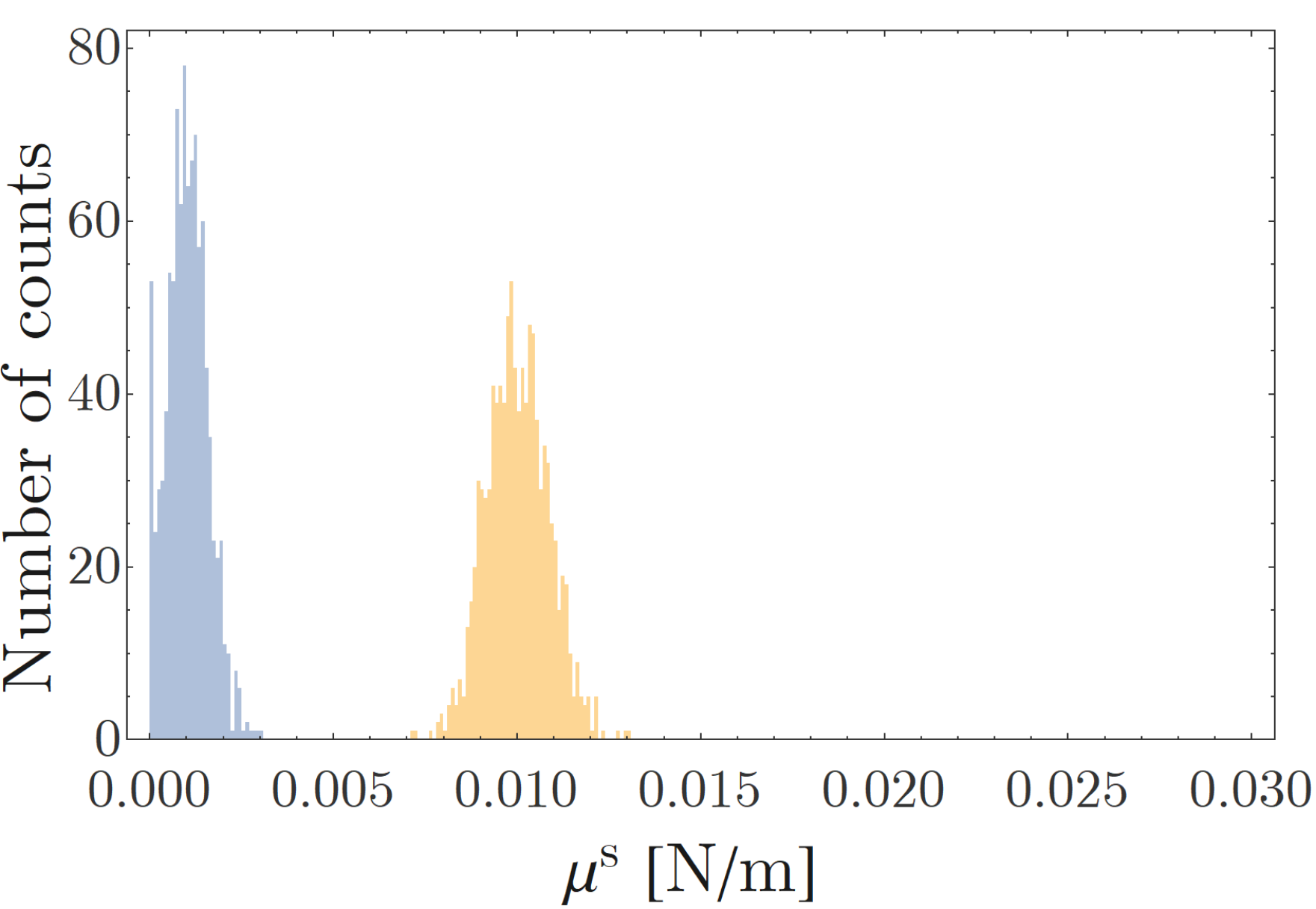}};
      \node[rectangle,
    draw = mathematica_blue,
    line width = 1.0mm,
    minimum width = 2.8cm, 
    minimum height = 2.6cm] (r) at (1.2,0.9) {};
      \node[] at (1.2,1.8) {$\mu^0\ll(\Upsilon^0,\kappa^0)$};
      \node[] at (1.2,1.) {$\Upsilon^0=10\,\text{mN/m}$};
      \node[] at (1.2,0.5) {$\mu^0=1\,\text{mN/m}$};
      \node[] at (1.2,0) {$\kappa^0=20\,\text{mN/m}$};
  \end{tikzpicture}  
\end{subfigure}
\begin{subfigure}{.49\textwidth}
  \centering
  \begin{tikzpicture}
      \node (histo_4) at (0,0) {\includegraphics[width=0.9\textwidth]{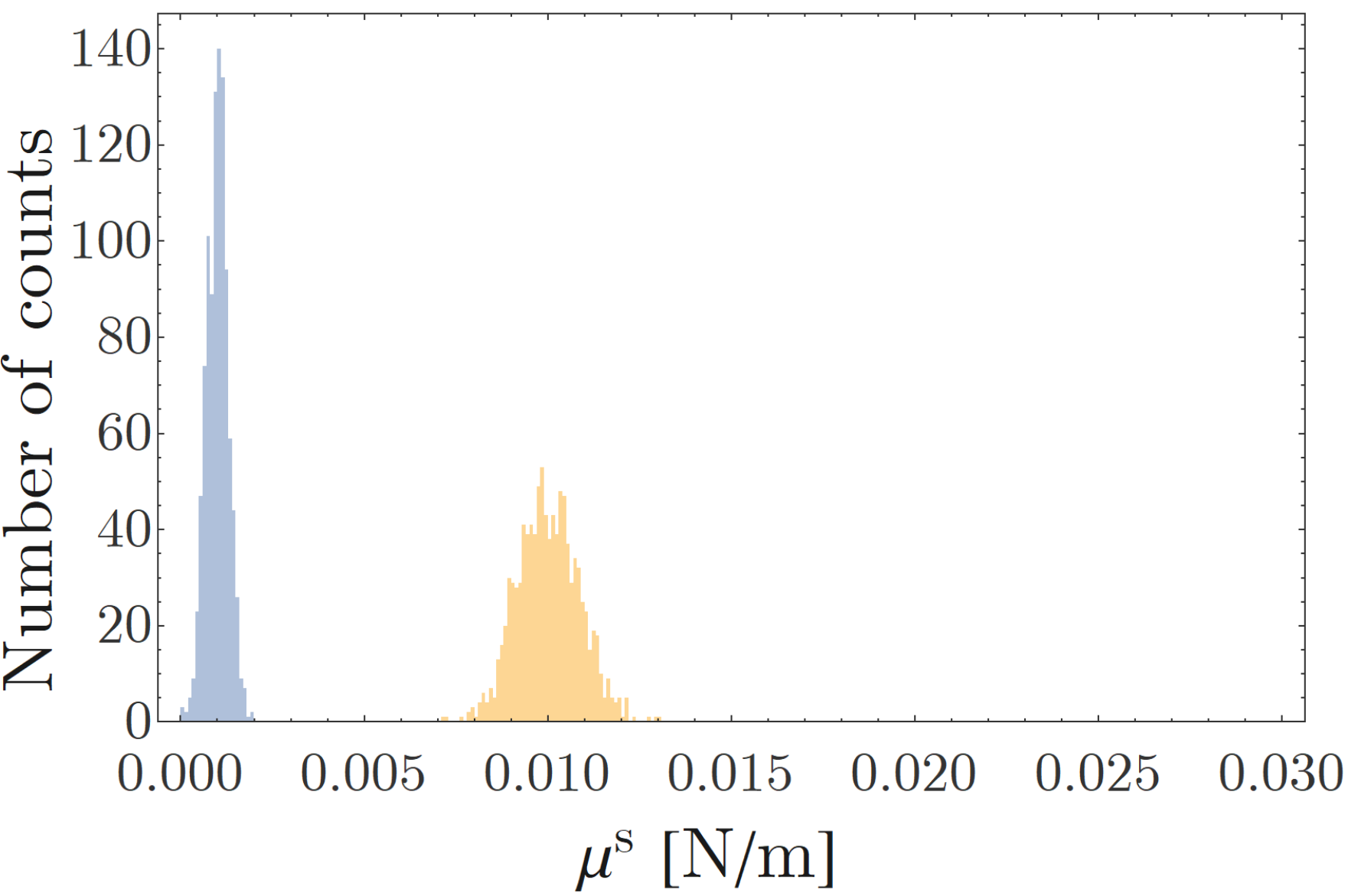}};
      \node[rectangle,
    draw = mathematica_blue,
    line width = 1.0mm,
    minimum width = 2.8cm, 
    minimum height = 2.6cm] (r) at (1.2,0.9) {};
    \node[] at (1.2,1.8) {$(\Upsilon^0,\mu^0)\ll\kappa^0$};
    \node[] at (1.2,1) {$\Upsilon^0=1\,\text{mN/m}$};
    \node[] at (1.2,0.5) {$\mu^0=1\,\text{mN/m}$};
    \node[] at (1.2,0) {$\kappa^0=20\,\text{mN/m}$};
  \end{tikzpicture} 
\end{subfigure}
\caption{Fitting sensitivities of $\mu^s$ for varying parameter ranges (blue) in comparison to results for $\Upsilon^0\sim(\mu^s,\kappa^s)$ (yellow), both at high levels of experimental error $\hat{\sigma}\sim 10\%$.}
\label{fig:Histo_Appendix_2}
\end{figure}

Figures~\ref{fig:Histo_Appendix} and~\ref{fig:Histo_Appendix_2} show obtained spectra for different cases of parameter ranges (blue) in comparison to the spectrum obtained from surface parameters on the same order of magnitude (yellow). 
As shown, both $\Upsilon^0$ and $\mu^s$ may be reliably recovered, whereby we find a decrease in the width of spectra in all cases. 

\end{document}